\documentstyle[pra,aps,epsbox]{revtex}

\tighten

\begin{document}

\draft
\title{
Time evolution of condensed state of interacting bosons 
with reduced number fluctuation in a 
leaky box
}
\author{
Akira Shimizu\cite{shmz} and Jun-ichi Inoue\cite{inoue}
}
\address{
Department of Basic Science, University of Tokyo, 
Komaba, Meguro-ku, Tokyo 153-8902, Japan
\\
and Core Research for Evolutional Science and Technology (CREST), 
JST
}
\maketitle
\begin{abstract}
We study the nonequilibrium time evolution of the Bose-Einstein condensate 
of interacting bosons confined in a leaky box, 
when its number fluctuation is initially ($t=0$) suppressed.
We take account of quantum fluctuations of all modes, 
including ${\bf k}={\bf 0}$, of the bosons.
As the wavefunction of the 
ground state that has a definite number $N$ of interacting bosons,
we use a variational form ${| N, {\bf y} \rangle}$, which
is obtained by operating a unitary operator $e^{i G({\bf y})}$ on the 
number state of free bosons.
Using  $e^{i G({\bf y})}$, 
we identify 
a ``natural coordinate'' ${\hat b_0}$ 
of the interacting bosons, 
by which many physical properties can be simply described.
%
The ${| N, {\bf y} \rangle}$ can be represented simply as 
a number state of $\hat b_0$, we thus call it
the ``number state of interacting bosons'' (NSIB).
To simulate real systems, for which if one fixes $N$ at $t=0$
$N$ will fluctuate at later times because of 
a finite probability of
exchanging bosons between the box and the environment,
we evaluate 
the time evolution of the reduced density operator 
$
\hat \rho (t) 
$
of the bosons in the box as a function of 
the leakage flux $J$.
We concentrate on the most interesting and nontrivial
time stage, {\it i.e.},  
the {\em early time stage} for
which $Jt \ll N$, 
much earlier than the time when
the system approaches the equilibrium state.
It is shown that 
the time evolution can be 
described very simply 
as the evolution from a single NSIB at $t < 0$,
into a classical mixture, with a time dependent distribution, of 
NSIBs of various values of $N$ at $t > 0$.
Using ${\hat b_0}$, 
we successfully define the cosine and sine operators 
{\em for interacting many bosons}, 
by which we can analyze the phase fluctuation
in a fully quantum-mechanical manner.
We define 
a new state 
${| \xi, N, {\bf y} \rangle}$
called
the ``number-phase squeezed state of interacting bosons'' (NPIB), 
which is characterized by a complex parameter $\xi$.
It is shown that $\hat \rho(t)$ for $t>0$ can be rewritten as 
the phase-randomized mixture (PRM) of NPIBs.
Among many possible representations of $\hat \rho(t)$, 
this representation is particularly convenient
for analyzing the phase fluctuations and the order parameter.
We study the order parameter according to 
a few typical definitions, as well as their time 
evolution.
It is shown that 
the off-diagonal long-range order (ODLRO) 
does not distinguish the NSIB and NPIB. 
Hence, the order parameter $\Xi$ defined
from ODLRO, 
does not distinguish them, either.
On the other hand, 
the other order parameter $\Psi$, defined as the expectation value 
of the boson operator $\hat \psi$, 
has different values among these states. 
In particular, for each element of the PRM of NPIBs, we show that 
$\Psi$ evolves from 
zero to a finite value very quickly.
Namely, after the leakage of only two or three bosons,
each element acquires a 
full, stable and definite (non-fluctuating) value of $\Psi$.
%
\end{abstract}


\section{Introduction and Summary}
\label{intro}

The Bose-Einstein condensation (BEC) has been observed 
in various systems \cite{bec93},
including the liquid Helium \cite{He4}, 
excitons in photo excited semiconductors \cite{BECexciton}, 
and atoms trapped by laser beams \cite{BECatom1,BECatom2,BECatom3}.
Although BEC was originally discussed 
for free bosons,
a condensate of free bosons does not have 
the superfluidity \cite{bec93}, 
hence many-body interactions are essential to interesting behaviors
of condensates.
The condensed state of interacting bosons in a box of finite volume $V$
is conventionally taken as 
the state in the Bogoliubov approximation,
which we denote by $| \alpha_0, {\bf y}^{cl} \rangle^{cl}$
[see Refs.\ \cite{Noz,popov} and Eq.\ (\ref{GScl}) below].
In this state, 
the boson number $N$ has finite fluctuation,
whose magnitude is 
$\langle \delta N^2 \rangle \gtrsim \langle N \rangle$
[Eq.\ (\ref{dN2cl})]. 
This fluctuation is non-negligible in small systems, 
such as Helium atoms in a micro bubble \cite{He4bubble} and 
laser-trapped atoms \cite{BECatom1,BECatom2,BECatom3}, where 
$\langle N \rangle$ is typically $10^3 - 10^6$, 
and thus 
$
\sqrt{\langle \delta N^2 \rangle} / \langle N \rangle 
=
3 - 0.1 \%
$.
This means that 
in such systems
if one fixes $N$ with the accuracy better than $3 - 0.1 \%$,
then $\langle \delta N^2 \rangle < \langle N \rangle$, thus 
the state $| \alpha_0, {\bf y}^{cl} \rangle^{cl}$ is forbidden,
and another state should be realized.
In most real systems, there is a finite probability of
exchanging bosons between the box and the environment.
Hence, if one fixes $N$ at some time, 
$N$ will fluctuate at later times.
Namely, the boson state undergoes a nonequilibrium 
time evolution when its number fluctuation is initially suppressed.
The purpose of this paper is to investigate
the time evolution of the interacting bosons in such a case,
and to discuss how an order parameter is developed.

We first review and discuss the case where the box is closed 
and the boson number $N$ is exactly fixed (section \ref{sec_pre}).
The ground-state wavefunction of such a case  
may be obtained 
by the superposition of 
Bogoliubov's solution $| \alpha_0, {\bf y}^{cl} \rangle^{cl}$
over various values of the phase of $\alpha_0$
[Eq.\ (\ref{inverse_tr_cl})]. 
The resulting state 
$| N, {\bf y}^{cl} \rangle^{cl}$
 has the same energy as 
$| \alpha_0, {\bf y}^{cl} \rangle^{cl}$
because of the degeneracy with respect to the phase of $\alpha_0$.
(This degeneracy leads to the symmetry breaking.)
However, such an expression is not convenient for the 
analysis of physical properties.
To find the ground state 
in the form that is convenient for analyzing 
physical properties, 
we derive an effective Hamiltonian 
$\hat H$ [Eq.\ (\ref{H})], which includes quantum fluctuations of all modes 
including ${\bf k} = 0$, 
 from the full Hamiltonian
$\hat H_B$ of interacting bosons.
Here, we neglect effects due to spatial inhomogeneity
of the boson states in the box, 
because we are not interested in such effects here, 
and also because we expect that main physics of the nonequilibrium evolution 
of our interest would not be affected by such effects.
A renormalization constant $Z_g$ appears in $\hat H$.
Although $Z_g$ formally diverges [Eq.\ (\ref{Zg})], 
the divergence is successfully 
renormalized, i.e., the final results are independent of $Z_g$
and finite. 
As the ground state of $\hat H$, 
we use a 
variational form ${| N, {\bf y} \rangle}$, 
which is similar to that of Girardeau and Arnowitt \cite{GA}.
This form takes a compact form 
[Eq.\ (\ref{GS in terms of a})]: it is obtained 
by operating a simple unitary operator $e^{i G({\bf y})}$ on the 
$N$-particle state of free bosons, 
where $G({\bf y})$ is a simple bi-quadratic function of the bare 
operators [Eq.\ (\ref{G})].
This state has the same energy as 
$| \alpha_0, {\bf y}^{cl} \rangle^{cl}$, or, equivalently, 
$| N, {\bf y}^{cl} \rangle^{cl}$. 
(Precisely speaking, they have the same energy 
density in the macroscopic limit, i.e., when $V \to \infty$ while 
keeping the density $n$ finite.)

Using the unitary operator $e^{i G({\bf y})}$, 
we then identify 
a ``natural coordinate'' ${\hat b_0}$ 
[Eq.\ (\ref{bz})] of the interacting bosons, 
by which many physical properties 
can be simply described (section \ref{sec_natural}).
Unlike the quasi-particle operators obtained by 
the Bogoliubov transformation,  
${\hat b_0}$ is a nonlinear function of bare operators.
Moreover, the Hamiltonian is {\em not} diagonal with respect to 
${\hat b_0}$.
Such a nonlinear operator, however, 
describes the physics very simply.
For example, 
${| N, {\bf y} \rangle}$ is simply represented as 
a number state of $\hat b_0$.
We thus call ${| N, {\bf y} \rangle}$ the ``number state of
interacting bosons'' (NSIB).
We can also define, through ${\hat b_0}$, the cosine and sine operators 
for interacting bosons [see below].
Moreover, using ${\hat b_0}$, we decompose the boson field $\hat \psi$ 
into two parts [Eq.\ (\ref{decomposition of psi})]:
one behaves anomalously as $V \to \infty$
and the other is a normal part.
In the decomposition, 
the non-unity ($|Z| < 1$) renormalization constant $Z$ 
(which should not be confused with $Z_g$ appeared in the Hamiltonian) 
is correctly 
obtained.  
This decomposition formula turns out to be 
extremely useful in the following 
analyses.

Using these results, 
we study the nonequilibrium time evolution of interacting 
bosons in a leaky box (section \ref{sec_evolution}).
The time evolution is induced if one fixes $N$ at some time 
(thus the boson state at that time is 
the NSIB), 
because in most real systems there is a finite probability of
exchanging bosons between the box and the environment,
hence $N$ will fluctuate at later times.  
We simulate this situation by the following gedanken experiment:
At some time $t<0$ 
one confines {\em exactly} $N$ bosons in a box of 
volume $V$, 
and that at $t=0$ 
a small hole is made in the box, 
so that a small leakage flux $J$ of the bosons is induced.
We concentrate on the analysis of 
the most interesting and nontrivial
time stage; 
the {\em early time stage} for
which $Jt \ll N$, because it is clear that at later times 
the system approaches the equilibrium state.
We are interested in 
the reduced density operator of the bosons in the box;
$
\hat \rho (t) 
\equiv
{\rm Tr^E} [ \hat \rho^{total}(t) ]
$, 
where 
$\hat \rho^{total}(t)$ denotes the density operator of the total 
system, and 
${\rm Tr^E}$ the trace operation over environment's 
degrees of freedom.
We successfully evaluate the time evolution of 
$
\hat \rho (t) 
$
by a method which is equivalent to solving the master equation.
Our method gives a physical picture more clearly than 
the master-equation method.
We obtain 
$
\hat \rho (t) 
$
in a general form in which all the 
details of the box-environment interaction $\hat H^{ES}$ have been 
absorbed in the magnitude of the leakage flux $J$.
We show that 
the time evolution can be 
described very simply in terms of ${\hat b_0}$,
as the evolution from a single NSIB at $t < 0$,
into a classical mixture, with a time dependent distribution, of 
NSIBs of various values of $N$ at $t > 0$ 
[Eq.\ (\ref{rho Poisson})].

We then discuss the phase $\phi$ as a variable approximately conjugate to 
the number $N$ (section \ref{sec_number vs phase}).
To treat the quantum phase properly, we consider the sine 
and cosine operators, $\hat {\sin \phi}$ and $\hat {\cos \phi}$.
It is generally very difficult to define such operators
for interacting many-particle systems.
Fortunately, however, in terms of the natural coordinate $\hat b_0$
we successfully define 
$\hat {\sin \phi}$ and $\hat {\cos \phi}$ for interacting bosons, 
using which we can analyze the phase property 
in a fully quantum-mechanical manner.
We define a ``coherent state of interacting bosons'' 
(CSIB) [Eq.\ (\ref{cs})], 
which, unlike Bogoliubov's ground state
$| \alpha_0, {\bf y}^{cl} \rangle^{cl}$,
 {\em exactly} has the minimum value of the 
number-phase uncertainty product [Eq.\ (\ref{NPUP_CSIB})].
We also define a new state 
${| \xi, N, {\bf y} \rangle}$ [Eq.\ (\ref{NPIB})], 
which we call
the ``number-phase squeezed state of interacting bosons'' (NPIB),
which has a smaller phase fluctuation than the CSIB, while 
keeping the number-phase uncertainty product 
minimum [Eq.\ (\ref{NPUP_NPIB})].
We point out that 
$\hat \rho(t)$ for $t > 0$ 
can be represented as 
the phase-randomized mixture (PRM) of NPIBs.
Among many possible representations of $\hat \rho(t)$, 
this representation is particularly convenient
for analyzing the phase fluctuations and the order parameter.

We also discuss the action of the measurements (or, 
their equivalents) of $N$ and of $\phi$ 
(section \ref{sec_action of meas.}). 
The forms of $\hat \rho(t)$ after such measurements 
are discussed.
As an example of the phase measurement,
we discuss an interference experiment of
two condensates which are prepared independently in
two boxes.
It was already established for {\em non-interacting} bosons 
that the interference pattern is developed for {\em each}
experimental run (although the interference 
disappears in the {\em average} over many runs)
\cite{theory1,theory2,theory3}.
Using our formula for interacting bosons,
we show that the same conclusion is drawn very clearly and naturally 
for {\em interacting} bosons.

We finally consider the order parameter according to 
a few typical definitions, as well as their time 
evolution (section \ref{sec_OP}).
We show that 
the off-diagonal long-range order (ODLRO) 
does not distinguish 
NSIB, NPIB and CSIB.
Hence, the order parameter $\Xi$ defined
from ODLRO [Eq.\ (\ref{Xi})], 
does not distinguish them either.
On the other hand, 
the other order parameter $\Psi$, defined as the expectation value 
of the boson operator $\hat \psi$, 
has different values among these states. 
In particular, for each element of 
the PRM of NPIBs, we show that 
$\Psi$ evolves from 
zero to a finite value very quickly:
After the leakage of only two or three bosons,
each element acquires a
full, stable and definite (non-fluctuating) value of $\Psi$.

\section{Preliminaries}
\label{sec_pre}

\subsection{Phase transitions in finite systems}
\label{sec_PT}

We consider the phase transition of 
interacting bosons confined in a large, but {\em finite} box
of volume $V$.
Phase transitions are usually discussed in systems
with an {\em infinite} volume
(or, the $V \to \infty$ limit is taken at the 
end of the calculation), because 
infinite degrees of freedom 
are necessary in the relevant energy scale 
for {\em strict} transitions \cite{gold}.
In such a case, 
we must select a single physical Hilbert space among 
many possibilities, 
which selection corresponds to a strict phase transition. 
However, phase transitions do occur 
even in systems of finite $V$ 
in the sense that a single phase lasts longer than the time of observation 
if its linear dimension exceeds the correlation length 
at the temperature of interest 
\cite{gold}.
Hence, it is physically interesting and important to 
explore phase transitions in finite systems.
Because of the finiteness of $V$ 
(and the fact that the interaction potential $U$ is well-behaved),
the von Neumann's uniqueness theorem can be applied. 
This allows us to 
develop a theory in a unique Hilbert space.
However, since $V$ is large,
some quantities, which become anomalous in the 
limit of $V \to \infty$ due to a strict phase transition,
behave quasi anomalously.
In later sections, we shall identify such a quasi anomalous 
operator, and discuss how an order parameter is 
developed.

\subsection{Effective Hamiltonian}
\label{sec_H}

We start from the standard Hamiltonian for interacting bosons
confined in a large, but finite box of volume $V$:
\begin{equation}
\hat H_B
=
\int_V d^3 r 
\hat \psi^\dagger({\bf r}) 
\left(- {\hbar^2 \over 2m} \nabla^2 \right)
\hat \psi({\bf r})
+
{1 \over 2}
\int_V d^3 r 
\int_V d^3 r'
\hat \psi^\dagger({\bf r}) 
\hat \psi^\dagger({\bf r'}) 
U({\bf r} - {\bf r'})
\hat \psi({\bf r'})
\hat \psi({\bf r}).
\label{starting H}\end{equation}
Here, we neglect a confining potential of the box
because in the present work we are not interested in its effects
such as the spatial inhomogeneity of the boson states in the box, 
and also because we expect that main physics of the nonequilibrium evolution 
of our interest would not be affected by such effects.
(Mathematically, our model under the periodic boundary condition 
assumes bosons confined in a three-dimensional torus.)
The ${\bf r}$ dependence of the boson field
$\hat \psi({\bf r})$ (in the Schr\"odinger picture)
can be expanded in terms of plane waves as
\begin{equation}
\hat \psi({\bf r})
=
{1 \over \sqrt{V}} \hat a_0
+
\sum_{{\bf k} \neq {\bf 0}} {e^{i {\bf k} \cdot {\bf r}} \over \sqrt{V}} 
\hat a_{\bf k},
\label{psi in terms of a}\end{equation}
where
$\hat a_{\bf p}$ and $\hat a_{\bf p}^\dagger$
are called creation and annihilation operators, 
respectively, of bare bosons.
The total number of bosons is given by 
\begin{equation}
\hat N 
\equiv
\int_V d^3 r \hat \psi^\dagger (r) \hat \psi(r)
= 
\sum_{\bf k} \hat a_{\bf k}^\dagger \hat a_{\bf k}.
\end{equation}
We assume zero temperature, and consider the case
where the interaction is weak and repulsive [Eq.\ (\ref{dilute}) below],
and where the boson density $n$ is finite
(hence, since $V$ is large, $N \gg 1$):
\begin{equation}
n \equiv \langle N \rangle /V > 0.
\label{density}\end{equation}
In such a case, BEC occurs and  
typical matrix elements of 
$\hat a_0$, $\hat a_0^\dagger$ and $\hat N$ 
are huge, whereas those of
$\hat a_{\bf k}$ and $\hat a_{\bf k}^\dagger$ (with ${\bf k} \neq {\bf 0}$) are small. 
Taking up to the second order terms in these small quantities,
and using the identity,
$
\hat N 
= 
\hat a_0^\dagger \hat a_0 
+ 
\sum_{{\bf k} \neq {\bf 0}} \hat a_{\bf k}^\dagger \hat a_{\bf k},
$
we obtain the effective Hamiltonian $\hat H$ in the following form.
\begin{equation}
\hat H 
=
g (1+Z_g) {\hat N^2 \over 2 V}
- g (1+Z_g){1 \over 2 V} \hat a_0^\dagger \hat a_0 
+
\sum_{{\bf k} \neq {\bf 0}} 
\left( 
\epsilon_k^{(0)}
+ g \frac{\hat N}{V}
\right)
\hat a_{\bf k}^\dagger \hat a_{\bf k}
+
\left[
{g \over 2V} \hat a_0 \hat a_0 
\sum_{{\bf k} \neq {\bf 0}} \hat a_{\bf k}^\dagger \hat a_{-{\bf k}}^\dagger
+ {\rm h.c.}
\right].
\label{Horg}\end{equation}
Here, $\epsilon^{(0)}_k$ denotes the free-particle energy,
$ 
\epsilon^{(0)}_k
\equiv 
{\hbar^2 k^2 / 2 m},
$
and $g$ is an effective interaction constant
defined by 
\begin{equation}
g \equiv {4 \pi \hbar^2 a \over m}.
\end{equation}
Here, $a$ is the scattering length, 
and 
$Z_g$ is the first-order ``renormalization constant'' 
for the scattering amplitude \cite{LP};
\begin{equation}
Z_g 
\equiv
{g \over 2 V} \sum_{{\bf k} \neq {\bf 0}} {1 \over \epsilon^{(0)}_k}.
\label{Zg}\end{equation}
The formal divergence of the sum in Eq.\ (\ref{Zg}) does not matter because
the final results are independent of $Z_g$ \cite{LP}, 
hence the renormalization is successful.
We have assumed that 
\begin{equation}
0 < n a^3 \ll 1,
\label{dilute}\end{equation}
under which the approximation $\hat H_B \approx \hat H$ is good.
We have confirmed by explicit calculations 
that the term
$
- \{ g (1+Z_g)/2 V \} \hat a_0^\dagger \hat a_0 
$
in Eq.\ (\ref{Horg}) gives only negligible contributions
 in the following analysis.
We thus drop it henceforth:
\begin{equation}
\hat H
=
g (1+Z_g) {\hat N^2 \over 2 V}
+
\sum_{{\bf k} \neq {\bf 0}} 
\left( 
\epsilon_k^{(0)}
+ g \frac{\hat N}{V}
\right)
\hat a_{\bf k}^\dagger \hat a_{\bf k}
+
\left[
{g \over 2V} \hat a_0 \hat a_0 
\sum_{{\bf k} \neq {\bf 0}} \hat a_{\bf k}^\dagger \hat a_{-{\bf k}}^\dagger
+ {\rm h.c.}
\right].
\label{H}\end{equation}
Since this $\hat H$ commutes with $\hat N$, 
we can in principle find its eigenstates for which $N$ is 
exactly fixed. 
In each subspace of fixed $N$, 
$\hat H$ is equivalent to
\begin{equation}
\hat H(N)
\equiv
{1 \over 2} g (1+Z_g) n^2 V
+
\sum_{{\bf k} \neq {\bf 0}} \epsilon'_k \hat a_{\bf k}^\dagger \hat a_{\bf k}
+
\left[
{g \over 2V} \hat a_0 \hat a_0 
\sum_{{\bf k} \neq {\bf 0}} \hat a_{\bf k}^\dagger \hat a_{-{\bf k}}^\dagger
+ {\rm h.c.}
\right],
\label{HN}\end{equation}
where 
\begin{equation}
\epsilon'_k
\equiv 
\epsilon_k^{(0)} + g n.
\label{ek}\end{equation}
Note that if we regard
$\hat a_0$ in $\hat H(N)$ 
as a classical complex number\cite{root} 
\begin{equation}
\hat a_0
\ \to \ 
e^{i \phi} \sqrt{N_0} \equiv \alpha_0,
\end{equation}
and $\hat a_0^\dagger$ as $\alpha_0^*$,  
we would then obtain 
the ``semiclassical'' Hamiltonian $\hat H^{cl}$ as 
\begin{equation}
\hat H^{cl}
=
{1 \over 2} g (1+Z_g) n^2 V
+
\sum_{{\bf k} \neq {\bf 0}} \epsilon'_k \hat a_{\bf k}^\dagger \hat a_{\bf k}
+
\left(
{1 \over 2} g n e^{2 i \phi}
\sum_{{\bf k} \neq {\bf 0}} \hat a_{\bf k}^\dagger \hat a_{-{\bf k}}^\dagger
+ {\rm h.c.}
\right),
\label{Hcl}\end{equation}
where 
we have replaced $N_0$ with $N$ in the last parenthesis 
because the replacement just gives correction which is 
of higher order in $g$.
This Hamiltonian can be diagonalized exactly (See, e.g., Ref.\ \cite{LP}
in which $\phi = 0$).
We shall utilize this fact later to find the ground state of $\hat H$.


\subsection{Known results for non-interacting bosons}

When $g=0$, the ground state of {\em free} bosons whose number $N$ 
is fixed, is simply a number state,
\begin{equation}
| N \rangle 
\equiv
{1 \over \sqrt{N!}} (\hat a_0^\dagger)^N | 0 \rangle,
\end{equation}
where $| 0 \rangle$ denotes 
the vacuum of the bare operators;
\begin{equation}
\hat a_{\bf k} | 0 \rangle =0 \quad \mbox{for all ${\bf k}$}.
\label{vac of a}\end{equation}
The energy of $| N \rangle$, 
\begin{equation}
E_{N}
=
N \epsilon_0^{(0)}
= 0,
\end{equation}
is degenerate with respect to $N$. Hence,
any superposition of $| N \rangle$ is also a ground state.
For example, the coherent state 
\begin{equation}
| \alpha \rangle 
\equiv
e^{-|\alpha|^2/2} 
\sum_{N =0}^\infty
\frac{\alpha^N}{\sqrt{N!}} | N \rangle
=
e^{-|\alpha|^2/2} e^{\alpha \hat a_0^\dagger} | 0 \rangle,
\end{equation}
where 
$ 
\alpha \equiv e^{i \phi} \sqrt{N},
$ 
is also a ground state that has the same expectation value of 
$\hat N$ as $|N \rangle$.
On the other hand, 
$| \alpha \rangle$ has a finite fluctuation of $N$,
\begin{equation}
\langle \delta N^2 \rangle
\equiv
\langle (\hat N - \langle \hat N \rangle )^2  \rangle
\nonumber\\
=
|\alpha|^2
\nonumber\\
=
\langle N \rangle,
\end{equation}
whereas $|N \rangle$ has a definite $N$.
The inverse transformation from 
$| \alpha \rangle$ to $|N \rangle$ can be accomplished as
\begin{equation}
| N \rangle
=
\int_{-\pi}^{\pi} \frac{d \phi}{2 \pi} 
| \alpha \rangle.
\label{inverse_tr_free}\end{equation}

\subsection{Known results for $\hat H^{cl}$}
\label{known1}

Neither 
$|N \rangle$ nor $| \alpha \rangle$ is an eigenstate when $g > 0$.
If we can regard 
$\hat a_0$ and $\hat a_0^\dagger$ as classical numbers 
$\alpha_0$ ($= e^{i \phi} \sqrt{N_0}$)
and $\alpha_0^*$, respectively,  
we can use $\hat H^{cl}$ as the Hamiltonian, and 
its ground state was given by Bogoliubov as \cite{bec93,Noz,popov}
\begin{equation}
| \alpha_0, {\bf y}^{cl} \rangle^{cl} 
\equiv
\exp \left[
\left( \alpha_0 \hat a_0^\dagger 
- {1 \over 2}
\sum_{{\bf q} \neq {\bf 0}} y_q^{cl *} \hat a_{\bf q}^\dagger \hat a_{-{\bf q}}^\dagger 
\right)- {\rm h.c.} 
\right]
| 0 \rangle,
\label{GScl}\end{equation}
where 
\begin{eqnarray}
y_q^{cl}
&=&
|y_q^{cl}|e^{-2 i \phi},
\label{ycl}\\
\cosh |y_q^{cl}| 
&=&
\sqrt{
\epsilon_q + \epsilon^{(0)}_q + gn
\over
2 \epsilon_q
},
\label{cosh y cl}\\ 
\sinh |y_q^{cl}| 
&=&
{
gn
\over 
\sqrt{2 \epsilon_q ( \epsilon_q + \epsilon^{(0)}_q + gn)}
}.
\label{sinh y cl}\end{eqnarray}
Here, $\epsilon_q$ is the quasi-particle energy,
\begin{equation}
\epsilon_q
\equiv
\sqrt{
\epsilon^{(0)}_q (\epsilon^{(0)}_q + 2 gn)
},
\end{equation}
whose dispersion is linear for
$\epsilon^{(0)}_q \ll gn$;
\begin{equation}
\epsilon_q
\approx
\sqrt{gn \over m} |q|.
\end{equation}
The ground state energy is calculated as \cite{LP}
\begin{eqnarray}
E_{\alpha, {\bf y}^{cl}}^{cl}
&=&
{1 \over 2} g n N
\left( 1 + {128 \over 15} \sqrt{n a^3 \over \pi} \right)
\label{Ecl}\\
&=&
{1 \over 2} g n N
+ O(g^{2.5}).
\end{eqnarray}
The absence of the $O(g^2)$ term in 
$E_{\alpha, {\bf y}^{cl}}^{cl}$ means that
the large (formally divergent because $Z_g \to \infty$)
positive energy $g Z_g n N/2$ of $\hat H^{cl}$ 
is canceled by a large negative term arising from 
the pair correlations.
The expectation value and variance of $\hat N$ for 
$| \alpha_0, {\bf y}^{cl} \rangle^{cl}$ are evaluated as
\begin{eqnarray}
\langle N \rangle^{cl}
&=&
N_0 + \sum_{{\bf q} \neq {\bf 0}} (\sinh |y_q^{cl}|)^2,
\label{Ncl}\\
\langle \delta N^2 \rangle^{cl}
&=&
\langle N \rangle^{cl}
+ 
\sum_{{\bf q} \neq {\bf 0}} (\sinh |y_q^{cl}|)^4
\label{dN2cl}.
\end{eqnarray}

\subsection{Ground state of a fixed number of bosons} 
\label{sec_GS}


In analogy to Eq.\ (\ref{inverse_tr_free}), 
it is possible to construct 
an approximate ground state of $\hat H(N)$
from $| \alpha_0, {\bf y}^{cl} \rangle^{cl}$ as 
\begin{equation}
| N, {\bf y}^{cl} \rangle^{cl}
\equiv
\int_{-\pi}^{\pi} \frac{d \phi}{2 \pi} 
| \alpha_0, {\bf y}^{cl} \rangle^{cl},
\label{inverse_tr_cl}\end{equation}
where $N \equiv \langle N \rangle^{cl}$, Eq.\ (\ref{Ncl}).
We can obtain an explicit form of 
$| N, {\bf y}^{cl} \rangle^{cl}$ in the form of
an infinite series expansion, 
by inserting Eq.\ (\ref{GScl}) into the right-hand side (rhs) of
Eq.\ (\ref{inverse_tr_cl}), expanding the exponential function, 
and performing the $\phi$ integral.
However, such an expression is not convenient for the 
analysis of physical properties.
Note that in many-particle physics, 
it is generally difficult to evaluate physical properties 
even if the wavefunction is known.
It is therefore essential to find the ground state 
in the form that is convenient for analyzing 
physical properties.

Several formulations were developed for the condensation of interacting
bosons with a fixed $N$.
Lifshitz and Pitaevskii \cite{LP} developed a {\em formal} discussion 
for the case of fixed $N$.
However, they did not treat $\hat a_0$ as an operator, hence 
$\hat N$ was {\em not} conserved. 
For example, $\hat b_p^\dagger |m,N \rangle$
(in their notations)
did not have exactly $N+1$ bosons.
To treat interacting bosons with fixed $N$ more accurately, 
one has to include quantum fluctuations of all modes 
(including ${\bf k}={\bf 0}$), by treating 
$\hat a_0$ as an operator.
Such treatment was developed, for example, by  
Girardeau and Arnowitt \cite{GA}, 
Gardiner \cite{gardiner}, and Castin and Dum \cite{castin}.
A variational form was proposed in Ref.\ \cite{GA} 
for the wavefunction of the ground state.
The variational form takes account of 
{\em four}-particle correlations in an elaborate manner, 
and is normalized exactly.
On the other hand, 
in Refs.\ \cite{gardiner} and \cite{castin} no explicit form 
was derived for the ground-state wavefunction.
[These references are more interested in excited states and 
the spatially inhomogeneous case, rather than 
the ground-state wavefunction.]
We here use a variational form, which is 
similar to that of Ref.\ \cite{GA}, 
%
as the ground state of $\hat H(N)$;
\begin{equation}
{| N, {\bf y} \rangle}
\equiv
e^{i G({\bf y})} {1 \over \sqrt{N!}} (\hat a_0^\dagger)^N | 0 \rangle.
\label{GS in terms of a}\end{equation}
Here, $\hat G({\bf y})$ is the hermite operator defined by 
\begin{equation}
\hat G({\bf y}) 
\equiv
{-i \over 2 nV} 
\hat a_0^\dagger \hat a_0^\dagger
\sum_{{\bf q} \neq {\bf 0}} |y_q^{cl}| \hat a_{\bf q} \hat a_{-{\bf q}}
+ {\rm h.c.},
\label{G}\end{equation}
where ${\bf y} \equiv \{y_q\}$ are a set of variational 
parameters, which are taken as
\begin{equation}
y_q = |y_q^{cl}|.
\label{yq}\end{equation}
Using the well-known formula for arbitrary operators $\hat A$ and $\hat B$,
\begin{equation}
e^{\hat A} {\hat B} e^{-\hat A}
=
{\hat B} + [{\hat A}, {\hat B}] 
+ {1 \over 2!} [{\hat A},[{\hat A}, {\hat B}]]
+ \cdots,
\label{formula}\end{equation}
we find from Eqs.\ (\ref{GS in terms of a})-(\ref{G}) that
\begin{equation}
E_{N, {\bf y}} \equiv 
{\langle N, {\bf y} |} \hat H {| N, {\bf y} \rangle}
=
{1 \over 2} g n N
+ o(g^2).
\label{E 2nd order}\end{equation}
where $o(g^2)$ denotes terms which tend to zero as $g \to 0$,
faster than $g^2$.
This demonstrates that 
the large (formally divergent because $Z_g \to \infty$) 
positive energy $g Z_g n N/2$ in $\hat H(N)$
is canceled by a large negative term arising from 
the {\em four-particle} correlations of 
the state of Eq.\ (\ref{GS in terms of a}).
Moreover, we can also show that 
in the macroscopic limit ($V \to \infty$ while keeping $n$ constant),
$E_{N, {\bf y}}$ becomes 
as low as $E_{\alpha, {\bf y}^{cl}}^{cl}$ of Eq.\ (\ref{Ecl}).
Therefore, 
the form of Eq.\ (\ref{GS in terms of a})
is a good approximation to the ground state.
Note that ${| N, {\bf y} \rangle}$ is an eigenstate of $\hat N$;
\begin{equation}
\hat N {| N, {\bf y} \rangle}
=
N {| N, {\bf y} \rangle},
\end{equation}
hence ${| N, {\bf y} \rangle}$ 
is an (approximate) groundstate of $\hat H$;
\begin{equation}
\hat H {| N, {\bf y} \rangle}
=
\hat H(N) {| N, {\bf y} \rangle}.
\end{equation}
Note also that 
${| N, {\bf y} \rangle}$ is exactly normalized to unity 
because
$e^{i G({\bf y})}$ in Eq.\ (\ref{GS in terms of a}) is a
unitary transformation (although it becomes non-unitary in 
the limit of $V \to \infty$).

We should make a remark here.
In the case of 
$| \alpha, {\bf y}^{cl} \rangle^{cl}$ discussed in subsection 
\ref{known1}, 
$E_{\alpha, {\bf y}^{cl}}^{cl}$ becomes low enough only for the specific 
choice of the phase of $y_q^{cl}$ [Eq.\ (\ref{ycl})].
This phase relation is sometimes called 
the ``phase locking'' \cite{Noz}.
From this viewpoint, 
it is sometimes argued \cite{Noz,Leg} that 
``having a definite phase'' and the ``phase locking'' 
are necessary to achieve a low energy.
However, such a statement is rather misleading:
In our case, the phase locking corresponds to the fact that $y_q$'s are
real and positive.
On the other hand, 
our ground state ${| N, {\bf y} \rangle}$ has no fluctuation in $N$, 
hence {\it has no definite phase} \cite{phase} 
because of the number-phase uncertainty relation,
\begin{equation}
\langle \delta N^2 \rangle 
\langle \delta \phi^2 \rangle
\gtrsim
1/4.
\label{NPUR}\end{equation}
Nevertheless, the energy of ${| N, {\bf y} \rangle}$ is as low as 
$E_{\alpha, {\bf y}^{cl}}^{cl}$. 
That is, 
``having a definite phase'' is {\it not} necessary to 
achieve the ground-state energy, and thus
the term ``phase locking'' should be taken carefully.

\subsection{Ground state of $N - \Delta N$ bosons} 

The ground state of $N - \Delta N$ bosons
is given by 
Eqs.\ (\ref{GS in terms of a}) and (\ref{G}) in which 
$N$ is all replaced with $N - \Delta N$.
However, we are interested in the case where 
[see section \ref{sec_evolution}]
\begin{equation}
|\Delta N| \ll N.
\end{equation}
In this case, $\hat G({\bf y})$ and ${\bf y}$ 
(which are functions of the density of bosons) of $N - \Delta N$ bosons are
almost identical to those of $N$ bosons
because $(N - \Delta N)/V \approx N/V = n$.
Therefore, we can simplify the calculation by using 
the same $\hat G({\bf y})$ and ${\bf y}$ for all $\Delta N$.
That is, we take
\begin{equation}
{| N - \Delta N, {\bf y} \rangle}
=
e^{i G({\bf y})} {1 \over \sqrt{(N - \Delta N) !}} 
(\hat a_0^\dagger)^{N - \Delta N} 
| 0 \rangle,
\label{GS of N'bosons}\end{equation}
where $\hat G({\bf y})$ and ${\bf y}$ are those of $N$ bosons.
Despite the approximation, 
this state is exactly normalized and 
has exactly $N - \Delta N$ bosons;
\begin{equation}
\hat N {| N - \Delta N, {\bf y} \rangle}
=
(N - \Delta N) {| N - \Delta N, {\bf y} \rangle}.
\end{equation}

\section{Natural coordinate}
\label{sec_natural}

\subsection{Nonlinear Bogoliubov transformation}

Since we assume that $V$ is finite, 
$e^{i G({\bf y})}$ is a  
unitary operator (which, however, becomes non-unitary in 
the limit of $V \to \infty$).
Utilizing this fact, 
we define new boson operators $\hat b_{\bf k}$ by
\begin{equation}
\hat b_{\bf k} 
\equiv 
e^{i \hat G({\bf y})} \hat a_{\bf k} e^{-i \hat G({\bf y})}.
\label{b}\end{equation}
This operator satisfies the same commutation relations as
$\hat a_{\bf k}$;
\begin{equation}
[\hat b_{\bf p}, \hat b_{\bf q}^\dagger] = \delta_{{\bf p}, {\bf q}},
\quad
[\hat b_{\bf p}, \hat b_{\bf q}] = 
[\hat b_{\bf p}^\dagger, \hat b_{\bf q}^\dagger] = 0.
\end{equation}
Note that these relations are exact, 
in contrast to the operator 
$b_{\bf k}$ (${\bf k} \neq {\bf 0}$) of Ref.\ \cite{gardiner}.
Owing to the exact commutation relations,
we can define 
the vacuum of $\hat b_{\bf k}$'s by
\begin{equation}
\hat b_{\bf k} {| 0, {\bf y} \rangle}
=0 
\quad \mbox{for all ${\bf k}$}.
\label{vac of b}\end{equation}
From Eqs.\ (\ref{vac of a}), (\ref{G}) and (\ref{b}), 
we have \cite{note_vac}
\begin{equation}
{| 0, {\bf y} \rangle}
=
e^{i \hat G({\bf y})} | 0 \rangle.
\end{equation}

The transformation (\ref{b}) somewhat resembles the Bogoliubov 
transformation which diagonalizes $\hat H^{cl}$ \cite{LP}.
However, in contrast to Bogoliubov's quasi-particles 
(whose total number differs from $\hat N$ as an operator), 
the total number operator of $\hat b_{\bf k}$'s is 
identical to that of $\hat a_{\bf k}$'s because 
$[\hat N, \hat G({\bf y})] = 0$;
\begin{equation}
\sum_{\bf k} \hat b_{\bf k}^\dagger \hat b_{\bf k}
=
e^{i \hat G({\bf y})} \hat N e^{-i \hat G({\bf y})}
=
\hat N.
\end{equation}
This property is very useful in the following analyses.
On the other hand, 
the transformation (\ref{b}) is much more complicated than the Bogoliubov 
transformation:
The latter is 
a {\it linear} transformation connecting the bare operators with 
quasi-particle operators, whereas the former 
is a {\it nonlinear} transformation between 
the bare operators and the new boson operators.
For example, 
${\hat b_0}$ defined by 
\begin{equation}
{\hat b_0}
\equiv 
e^{i \hat G({\bf y})} \hat a_0 e^{-i \hat G({\bf y})}
\label{bz}\end{equation}
is a rather complicated, 
nonlinear function of  
$\hat a_{\bf k}$'s and $\hat a_{\bf k}^\dagger$'s 
(of various ${\bf k}$'s including ${\bf k} = {\bf 0}$), 
as can be seen using Eq.\ (\ref{formula}).
Such a nonlinear operator $\hat b_0$, however, describes the physics quite
simply.
Namely, we show that 
${\hat b_0}$ (and ${\hat b_0^\dagger}$) is a 
``natural coordinate'' 
\cite{hermitian} of interacting bosons
in the sense that many physical properties can be simply described.
(It is crucial to find such a coordinate for the analysis 
of many-particle systems, because generally the knowledge of 
the wavefunction is not sufficient to perform 
the analysis.)
For example, 
from Eqs.\ (\ref{GS in terms of a}), (\ref{b}) and (\ref{vac of b}), 
we find that in terms of ${\hat b_0}$ the ground state 
${| N, {\bf y} \rangle}$ is simply a number state;
\begin{equation}
{| N, {\bf y} \rangle}
=
{1 \over \sqrt{N !}} ({\hat b_0}^\dagger)^N {| 0, {\bf y} \rangle}.
\label{GS}\end{equation}
In particular, 
\begin{equation}
\hat b_0 {| N, {\bf y} \rangle}
=
\sqrt{N} {| N-1, {\bf y} \rangle}.
\label{app_b0}\end{equation}
Since $\hat b_{\bf k}$'s of ${\bf k} \neq {\bf 0}$ 
commute with $\hat b_0$, 
we also find that
\begin{equation}
\hat b_{\bf k} {| N, {\bf y} \rangle}
= 0
\quad \mbox{for all ${\bf k} \neq {\bf 0}$}.
\label{GS is vac of bk}\end{equation}
Therefore, in terms of the new boson operators 
the ground state of the interacting bosons 
can be simply viewed as a {\em single-mode} 
(${\bf k} = {\bf 0}$) number state.
Note, however, that 
this does not mean that $\hat H$ were 
bilinear and diagonal with respect to
$\hat b_0$. 
In fact, if it were the case then 
the energy $E_{N, {\bf y}}$ would be linear in $N$, in contradiction 
to Eq.\ (\ref{E 2nd order}) which shows $E_{N, {\bf y}} \propto N^2$
(recall that $n = N/V$). 

The usefulness of ${\hat b_0}$ are strongly suggested by 
Eqs.\ (\ref{GS}) and (\ref{GS is vac of bk}).
We will show in the following 
discussions that this is indeed the case.

\subsection{Decomposition of $\hat \psi$}

Some matrix elements of $\hat b_0$ become anomalously large, 
among the ground (and excited) states of different $N$.
For example, 
\begin{equation}
{\langle N - 1, {\bf y} |} {\hat b_0} {| N, {\bf y} \rangle} = \sqrt{N} = \sqrt{nV}.
\label{anomalous}\end{equation}
This
indicates that in the $V \to \infty$ limit 
(while keeping the density $n$ finite)
${\hat b_0}$ does not 
remain an annihilation operator of 
the physical Hilbert space,
signaling that a strict phase transition should occur as $V \to \infty$.
Since this anomaly should have important effects even for a finite $V$, 
it is appropriate to separate $\hat b_0$ from the other terms 
of $\hat \psi$. 
That is, 
we decompose the boson field in a finite system as 
({\it cf.} Eq.\ (\ref{psi in terms of a}))
\begin{equation}
\hat \psi 
=
{Z^{1/2} \over \sqrt{V}} \hat b_0 + \hat \psi',
\label{decomposition of psi}\end{equation}
where $Z$ is a complex renormalization constant.
Since we have specified nothing about $\hat \psi'$ at this stage, 
this decomposition is always possible and $Z$ is arbitrary.
Following Ref.\ \cite{LP}, we define 
the ``wavefunction of the condensate'' $\Xi$ by
\begin{equation}
\Xi 
\equiv
{\langle N - 1, {\bf y} |} \hat \psi {| N, {\bf y} \rangle}.
\label{wf of condensate}\end{equation}
Since $\Xi$ is independent of ${\bf r}$ (because 
both ${| N - 1, {\bf y} \rangle}$ and ${| N, {\bf y} \rangle}$ 
have the translational symmetry), 
we can take $Z$ as 
\begin{equation}
\Xi
=
Z^{1/2} {\langle N - 1, {\bf y} |} \hat b_0 {| N, {\bf y} \rangle} /\sqrt{V}.
\label{takeZ}\end{equation}
That is, from Eq.\ (\ref{GS}), 
\begin{equation}
Z^{1/2} = 
\Xi / \sqrt{n}.
\label{Z}\end{equation}
Then, by taking the matrix element of 
Eq.\ (\ref{decomposition of psi}) between ${| N - \Delta N, {\bf y} \rangle}$ and ${| N, {\bf y} \rangle}$, 
we find 
\begin{equation}
{\langle N - \Delta N, {\bf y} |}  \hat \psi' {| N, {\bf y} \rangle}
=
0
\quad \mbox{for $^\forall \Delta N$ ($|\Delta N| \ll N$)}.
\label{property of psi'}\end{equation}
We now define two number operators by
\begin{eqnarray}
\hat N' 
&\equiv& 
\int d^3 r \hat \psi^{\prime \dagger} (r) \hat \psi'(r),
\\
\hat N_0 
&\equiv& 
\hat N - \hat N'.
\label{N0}\end{eqnarray}
Then, from Eqs.\ (\ref{GS}), (\ref{decomposition of psi}),   
(\ref{Z}), and (\ref{property of psi'}), we find
\begin{equation}
{\langle N, {\bf y} |} \hat N {| N, {\bf y} \rangle} 
=
V |\Xi|^2 + {\langle N, {\bf y} |} \hat N' {| N, {\bf y} \rangle}.
\end{equation}
Hence, from Eq.\ (\ref{N0}), 
\begin{equation}
{\langle N, {\bf y} |} \hat N_0 {| N, {\bf y} \rangle} 
=
V |\Xi|^2,
\label{N0 vs Xi}\end{equation}
which may be interpreted as the 
``number of condensate particles'' \cite{LP}. 
That is, 
in agreement with the standard result \cite{LP}, 
$|\Xi|^2$ is the density of 
the condensate particles;
\begin{equation}
|\Xi|^2 = \langle N_0 \rangle / V \equiv n_0,
\end{equation}
where
we have denoted the expectation value simply by $\langle \cdots \rangle$. 
We can therefore write $\Xi$ as
\begin{equation}
\Xi = \sqrt{n_0} e^{i \varphi}.
\label{Xi_n0}\end{equation}
We thus find the formula for the decomposition of $\hat \psi$ as
\begin{equation}
\hat \psi
=
e^{i \varphi} \sqrt{n_0 \over n V} \hat b_0
+
\hat \psi',
\label{formula of decomposition}\end{equation}
which is extremely useful in the following analysis.
Note that we have obtained the finite renormalization; 
\begin{equation}
|Z| = n_0 / n < 1.
\end{equation}

\subsection{Relation to the previous work}

Lifshitz and Pitaevskii \cite{LP} {\em introduced} an operator $\hat \Xi$ 
that transforms an eigenstate with $N$ bosons into 
the corresponding eigenstate with $N-1$ bosons,
{\em without} giving an explicit form of $\hat \Xi$.
(They defined $\hat \Xi$ through its matrix elements 
between eigenstates with different values of $N$.
However, as mentioned in section \ref{sec_GS}, 
they did not give the forms of the eigenstates of fixed $N$.)
They decomposed $\hat \psi$ as (Eq.\ (26.4) of Ref.\ \cite{LP})
\begin{equation}
\hat \psi 
=
\hat \Xi + \hat \psi'.
\label{decomposition of psi by LP}\end{equation}
In the present paper, 
from Eqs.\ (\ref{bz}) and (\ref{formula of decomposition}), 
we obtain the {\em explicit} expression for $\hat \Xi$ as
\begin{equation}
\hat \Xi 
= e^{i \varphi} \sqrt{n_0 \over n V} \hat b_0
= e^{i \varphi} \sqrt{n_0 \over n V} 
e^{i \hat G({\bf y})} \hat a_0 e^{-i \hat G({\bf y})}
\end{equation}
From Eqs.\ (\ref{app_b0}) and (\ref{Xi_n0}), we confirm that 
\begin{equation}
\hat \Xi {| N, {\bf y} \rangle}
=
\Xi {| N-1, {\bf y} \rangle},
\label{app_Xi}\end{equation}
which was {\em assumed} in Ref.\ \cite{LP}.
The operator $\hat \Xi$ characterizes the condensation by having 
a finite matrix element \cite{LP}.
In the following we will reveal a striking property of
$\hat \Xi$ (or, equivalently, $\hat b_0$); it   
is a ``natural coordinate'' of interacting bosons.

On the other hand, 
other operators, which also characterize the condensation, 
were introduced in Refs.\ \cite{GA,gardiner,castin}.
Girardeau and Arnowitt \cite{GA}
defined $\hat \beta_0 \equiv \hat a_0 \hat N_0^{-1/2}$,
Gardiner \cite{gardiner} introduced 
$\hat A = \hat a_0 (\hat N_0/\hat N)^{1/2}$ (which is 
an operator form of Eqs.\ (9) and (10) of Ref.\ \cite{gardiner}),
and Castin and Dum \cite{castin} introduced 
$\hat a_{\Phi_{ex}}
\equiv
\int dr \Phi_{ex}^* (r,t) \hat \psi(r,t)
$.
These operators 
are totally different from $\hat \Xi$ or $\hat b_0$
because complicated many-particle correlations, 
which are included in $\hat \Xi$ and $\hat b_0$, 
are not included in $\hat \beta_0$, $\hat A$ and $\hat a_{\Phi_{ex}}$.
For example, $\hat a_{\Phi_{ex}}$ is a {\em linear} combination of 
{\em annihilation} operators of free bosons, 
whereas $\hat b_0$ is a {\em nonlinear} function of
{\em both} the annihilation and creation operators of free bosons.
As a result, 
in contrast to Eqs.\ (\ref{app_b0}) and (\ref{app_Xi}), 
application of either $\hat \beta_0$, $\hat A$ or $\hat a_{\Phi_{ex}}$
to the ground state of $N$ bosons does not yield
the ground state of $N-1$ bosons;
it yields an excited state which is not an eigenstate.
Moreover, they are not a natural coordinate of interacting bosons
in the sense explained in the following.
Therefore, we do not use these operators, 
although they would be useful in other problems.

\subsection{Low-lying excited states}

Excited states of a fixed number of interacting bosons 
were discussed in Refs.\ \cite{GA,gardiner,castin}.
In the present formulation, we may obtain low-lying excited states
by the application 
to ${| N, {\bf y} \rangle}$ of
functions of $\hat b_{\bf k}^\dagger$'s with ${\bf k} \neq {\bf 0}$.
However, 
since we do not need any  
explicit expressions of the excited states in the following analysis,
we do not seek for them in the present paper. 

\section{Time evolution of bosons in a leaky box}
\label{sec_evolution}

The time evolution of a condensate(s) in an open box(es) 
was discussed previously 
for the cases of {\em non-interacting} bosons 
in Refs.\ \cite{theory1,theory2,theory3} and 
for the case of {\em two-mode} interacting bosons in Ref.\ \cite{RW}.
In the present paper, 
using $\hat b_0$,
we study the case of infinite-mode interacting bosons.

\subsection{Gedanken experiment}

\begin{figure}[h]
\begin{center}
\epsfile{file=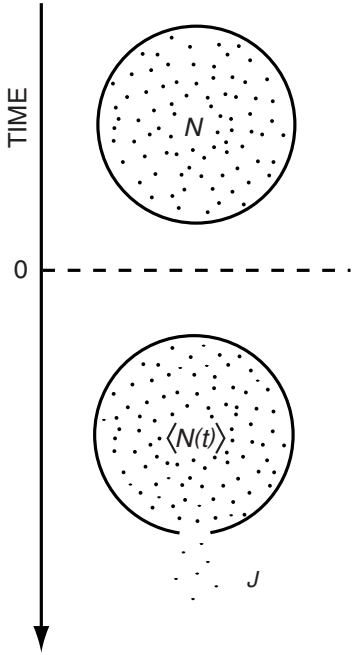,scale=0.4}
\end{center}
\caption{
Our gedanken experiment.
$N$ bosons are confined in a closed box for $t<0$.
At $t=0$ a small hole is made in the box,
so that a small leakage flux $J$ is induced, and
the expectation value $\langle N(t) \rangle$ of the number 
of bosons in the box decreases with time.
}
\label{gedanken}
\end{figure}

In most real systems, there is a finite probability of
exchanging bosons between the box and the environment.
Hence, even if one fixes $N$ at some time, 
$N$ will fluctuate at later times.
Namely, the boson state undergoes a nonequilibrium 
time evolution when its number fluctuation is initially suppressed.
To simulate this situation, 
we consider the following gedanken experiment [Fig.\ \ref{gedanken}].

Suppose that bosons are confined in a box
which is kept at zero temperature, 
and that the wall of the box is not permeable to 
the flow of the bosons, {\it i.e.}, 
the probability of boson's permeating through the wall 
within a time scale of our interest is negligible.
If one measures the number of the bosons
at a time $t=t_p$ ($< 0$), 
and if the box is kept closed until $t=0$, 
then the density operator $\hat \rho(t)$ of the bosons in the box for  
$t_p < t < 0$ is
\begin{equation}
\hat \rho(t) = {| N, {\bf y} \rangle} {\langle N, {\bf y} |}
\quad \mbox{for $t_p < t < 0$}.
\label{past rho}\end{equation}
Assume that this box is placed in a large room, which has 
no bosons initially. 
Suppose now that at $t=0$ he makes 
a small hole(s), or slightly lowers the potential of the wall
of the box, 
so that a small but finite flow $J$ of the bosons from 
the inside to the outside of the box 
becomes possible for $t \geq 0$.
We study the time evolution for $t \geq 0$ of the density operator 
$\hat \rho(t)$ of the bosons in the box.

The expectation value of $N$ will be a decreasing function
of $t$, which we denote ${\langle N(t) \rangle}$ 
(hence $\langle N(0) \rangle =N$).
It is obvious that 
as $t \to \infty$ the system approaches the equilibrium state. 
Therefore, we are most interested in 
{\em the early stage of the time evolution}, 
for which 
\begin{equation}
|N - {\langle N(t) \rangle}| \ll N.
\label{early}\end{equation}
Note that if $J$ were not small enough, 
then the state in the box would evolve into 
a nonequilibrium excited state.
The property of such a nonequilibrium state would 
depend strongly on details of the structures of the box 
and the hole or wall.
In the present paper, we are not interested in such structure-sensitive 
states.
Therefore, we assume that $J$ is small enough 
that only transitions between the ground states 
for different values of $N$ are possible.

\subsection{Total Hamiltonian}

Let ${\cal V}$ denote the volume of the room, 
which is much larger than the volume $V$ of the box;
\begin{equation}
{\cal V} \gg V.
\end{equation}
The total boson field $\hat \psi^{total}({\bf r})$
is defined on ${\cal V}$, 
\begin{equation}
\hat \psi^{total}({\bf r})
=
\hat \psi({\bf r}) + \hat \psi^E({\bf r}),
\end{equation}
where $\hat \psi({\bf r})$ is localized in the box, and 
$\hat \psi^E \equiv \hat \psi^{total} - \hat \psi$
is the boson field of ``environment.''
Then, the total Hamiltonian may take the following form:
\begin{equation}
\hat H^{total} 
=
\hat H
+
\hat H^E
+
\hat H^{ES},
\label{Htotal}\end{equation}
where $\hat H$ and 
$\hat H^E$ are the Hamiltonians 
of the $\hat \psi$ and $\hat \psi^E$ fields, 
respectively, when they are closed.
Small but finite amplitudes of scattering between 
the $\hat \psi$ and $\hat \psi^E$ fields
are caused by the residual Hamiltonian
$\hat H^{ES}$.
Under our assumption (\ref{dilute}), 
the probability of multi-particle collisions during 
the escape from the box is negligible.
Therefore, $\hat H^{ES}$ should take the following form:
\begin{equation}
\hat H^{ES}
=
\lambda
\int d^3 r 
\hat \psi^{E \dagger}({\bf r}) 
f({\bf r}) 
\hat \psi({\bf r})
+ {\rm h.c.},
\label{HES}\end{equation}
where $\lambda$ is a constant which has the dimension of energy, 
and $f({\bf r})$ is a dimension-less 
function which takes values of order unity 
when ${\bf r}$ is located in 
the boundary region between the box and the environment, 
and $f({\bf r})=0$ otherwise.
Although the value of $\lambda$ and the form of $f$ 
depend on the structures of the box and the hole or walls,
our final results (e.g., Eq.\ (
\ref{rho Poisson})) are independent of such details.

\subsection{Low-lying states of the total system}

We here list states of the total system which are 
relevant to the following analysis.
Since $\hat H^{ES}$ is weak, 
quasi eigenstates of the total system are well approximated by 
the products of eigenstates of the box and of the environment.
Recall that 
$| N - \langle N(t) \rangle| \ll N$ 
for the time interval of our interest, 
and that
$J$ is small enough so 
that only transitions between the ground states 
for different values of $N$ are possible.
Therefore, among many possible states of the box 
the relevant states are 
${| N - \Delta N, {\bf y} \rangle}$'s with $|\Delta N| \ll N$.
On the other hand, there are no 
bosons in the environment at $t<0$.
That is, the environment is initially in 
the vacuum, which we denote $| 0^{E} \rangle$. 
Hence, from Eq.\ (\ref{past rho}), 
the initial density operator of the total system is 
\begin{equation}
\hat \rho^{total}(t) 
=
{| N, {\bf y} \rangle} | 0^{E} \rangle \langle 0^{E} | {\langle N, {\bf y} |}
\quad (t<0) .
\label{initial rho total}\end{equation}
Bosons escape from the box into the environment for $t \geq 0$.
Since ${\cal V} \gg V$, the boson density of the environment is kept 
essentially zero,  
and BEC does not occur in the environment,
for the time period of Eq.\ (\ref{early}).
We can therefore take 
the simple number states
$| n^{E}_{\bf k}, n^{E}_{\bf k'}, \cdots \rangle$'s
of free bosons
as eigenstates of the environment, where
$n_{\bf k}$ denotes the number of bosons in mode ${\bf k}$.
For example, 
we shall write 
$| 1_{\bf k}^{E} \rangle$
to denote the environment state 
in which mode ${\bf k}$ are occupied by a single boson whereas 
the other modes are empty.
Therefore, the relevant states of the total system, 
{\it i.e.}, low-lying quasi eigenstates of $\hat H^{total}$, 
can be written as 
\begin{equation}
{| N - \Delta N, {\bf y} \rangle} | n^{E}_{\bf k}, n^{E}_{\bf k'}, \cdots \rangle
\quad (|\Delta N| \l N),
\label{relevant states}\end{equation}
where, since 
$\hat H^{total}$ conserves the total number of bosons,
\begin{equation}
\Delta N 
=
\sum_{\bf k} n^{E}_{\bf k}.
\label{conservation of total N}\end{equation}

\subsection{Time evolution in a short time interval $\Delta t$}

We are interested in the reduced density operator of 
bosons in the box for $t \ge 0$;
\begin{equation}
\hat \rho (t) 
\equiv
{\rm Tr^E} [ \hat \rho^{total}(t) ],
\label{reduced rho}\end{equation}
where ${\rm Tr^E}$ is the trace operation over the environment degrees of
freedom.
The expectation value of any observable $\hat Q$ in the box 
can be evaluated from $\hat \rho (t)$ as
\begin{eqnarray}
\langle Q(t) \rangle
&\equiv&
{\rm Tr^{total}}  [ \hat \rho^{total}(t) \hat Q ]
\nonumber\\
&=&
{\rm Tr}  [ \hat \rho(t) \hat Q ],
\end{eqnarray}
where 
${\rm Tr}$ denotes the trace operation over the degrees of
freedom in the box.
Equation (\ref{initial rho total}) yields 
\begin{eqnarray}
\hat \rho(0) 
&=&
{| N, {\bf y} \rangle} {\langle N, {\bf y} |} 
\label{rho at 0}\\
\langle Q(0) \rangle
&=&
{\langle N, {\bf y} |} \hat Q {| N, {\bf y} \rangle}.
\end{eqnarray}

Although we may evaluate $\hat \rho (t)$ 
by solving a master equation, 
we here present a different 
(but equivalent) method, 
by which the physical meaning can be seen clearly.
We begin with noting that a single action of 
$\hat H^{ES}$ of Eq.\ (\ref{HES}) can only change $(N, N^{E})$ 
by either $(-1,+1)$ or $(+1,-1)$, and that 
the latter change is impossible for the initial density operator
(\ref{initial rho total}).
Therefore, 
after a short time interval $\Delta t$ which satisfies $J \Delta t \ll 1$,
the state vector 
${| N, {\bf y} \rangle} | 0^{E} \rangle$ evolves 
into a state of the following form:
\begin{equation}
e^{-i E_{N, {\bf y}} \Delta t / \hbar}
{| N, {\bf y} \rangle} | 0^{E} \rangle
+
\sum_{\bf k} 
c_{\bf k}^{(1)} (\Delta t)
e^{-i (E_{N-1, {\bf y}} + \epsilon^{(0)}_k) \Delta t / \hbar}
{| N - 1, {\bf y} \rangle} | 1^{E}_{\bf k} \rangle 
+
O(\lambda^2),
\label{state after Dt}
\end{equation}
where
\begin{eqnarray}
c_{\bf k}^{(1)} (\Delta t)
&\equiv&
{1 \over i \hbar}
\int_0^{\Delta t} 
M_{\bf k} e^{i (\epsilon^{(0)}_k - \mu) \tau / \hbar} d \tau,
\\
M_{\bf k} 
&\equiv&
\langle 1_{\bf k}^{E}| {\langle N - 1, {\bf y} |}
\hat H^{ES} 
{| N, {\bf y} \rangle} | 0^{E} \rangle,
\\
\mu
&\equiv&
E_{N, {\bf y}} - E_{N-1, {\bf y}}.
\end{eqnarray}
Therefore, 
the reduced density operator 
is evaluated as
\begin{equation}
\hat \rho (\Delta t)
=
w(0;\Delta t) {| N, {\bf y} \rangle} {\langle N, {\bf y} |} 
+
w(1;\Delta t) {| N - 1, {\bf y} \rangle} {\langle N - 1, {\bf y} |} 
+
O(\lambda^3),
\label{rho at Delta t}\end{equation}
where 
\begin{eqnarray}
w(0;\Delta t)
&\equiv&
1 - w(1;\Delta t),
\label{w0}\\
w(1;\Delta t)
&\equiv&
\sum_{\bf k} |c_{\bf k}^{(1)}(\Delta t)|^2. 
\end{eqnarray}
Here, we have normalized $\hat \rho (\Delta t)$ to order $\lambda^2$ by 
Eq.\ (\ref{w0}).
We now take $\Delta t$ in such a way that
\begin{equation}
\hbar / E_c < \Delta t \ll 1/J, 
\label{range of Delta t}\end{equation}
where $E_c$ is the energy range of $\epsilon^{(0)}_{\bf k}$ in which 
$|M_{\bf k}|^2$ is finite and approximately constant.
Then, 
since ${\bf k}$ of the environment takes quasi-continuous values,
$w(1;\Delta t)$ becomes proportional to $\Delta t$:
\begin{equation}
w(1;\Delta t)
=
J \Delta t.
\end{equation}
To evaluate $J$, we calculate $|M_{\bf k}|^2$ using 
Eqs.\ (\ref{property of psi'}) and (\ref{HES}) as
\begin{eqnarray}
|M_{\bf k}|^2 
&=& 
\left|
\lambda
\int d^3 r 
f({\bf r}) 
\langle 1_{\bf k}^{E} | {\langle N - 1, {\bf y} |}
\hat \psi^{E \dagger}({\bf r}) 
\hat \Xi
{| N, {\bf y} \rangle} | 0^{E} \rangle
\right|^2
\nonumber\\
&=& 
N {n_0 \over n}
\left|
{\lambda \over \sqrt{V}}
\int d^3 r 
f({\bf r}) \varphi_{\bf k}^{E *}({\bf r})
\right|^2,
\end{eqnarray}
where $\varphi_{\bf k}^{E}({\bf r})$ is the mode function of mode 
${\bf k}$ of the environment. 
Regarding the volume dependence, 
$\varphi_{\bf k}^{E}$ behaves as $\sim 1/\sqrt{\cal V}$, 
whereas $f$ is localized in the boundary region, 
whose volume is denoted by $v$,  
of the box and the environment.
Therefore, 
\begin{equation}
|M_{\bf k}|^2 
\approx
{n_0 \over n}
|\lambda|^2
{v^2 \over V {\cal V}}.
\end{equation}
On the other hand, from Eq.\ (\ref{E 2nd order}), 
an escaping boson has 
the energy of $g n$.
The density of states of the environment at this energy is
\begin{equation}
{
{\cal V}
\over 
2 \pi^2 \hbar^3
}
\sqrt{m^3 g n \over 2}.
\end{equation}
Therefore, the leakage flux $J$ is estimated as
\begin{equation}
J
\approx
N
{n_0 \over n}
{v \over V}
{|\lambda|^2 v \over \hbar^4}
\sqrt{m^3 g n} 
=
\frac{
n_0 |\lambda|^2 v^2
}{
\hbar^4
} 
\sqrt{m^3 g n},
\label{J}
\end{equation}
where numerical factors of order unity have been absorbed in $\lambda$.
We observe that
$J$ is reduced by the factor $v/V$ ($\ll 1$), which means 
that the escape process is a ``surface effect'', 
i.e., it occurs only in the boundary region.
On the other hand, 
$J$ is enhanced by the factor $N$ ($\gg 1$).
This enhancement is typical to the boson condensation.

From our assumption of small $J$, 
the rhs of Eq.\ (\ref{J}) should be smaller than
the critical value $J_{cr}$
of the flux above which bosons in the box get excited \cite{Jcr}.
It is seen that this condition is satisfied when 
$v$ and/or $|\lambda|$ is small enough.

\subsection{Time evolution for $0 \leq Jt \ll N$}
\label{sec_later}

We have found in the previous subsection that
the reduced density operator $\hat \rho$ evolves 
from the pure state (\ref{rho at 0}) 
to the mixed state (\ref{rho at Delta t}) after 
a small time interval $\Delta t$ ($\ll 1/J$).
Since the latter 
is a classical mixture of two different states, 
${| N, {\bf y} \rangle} {\langle N, {\bf y} |}$ and ${| N - 1, {\bf y} \rangle} {\langle N - 1, {\bf y} |}$, we can 
separately solve the time evolution 
for each state.
For each state, further transitions occur in the 
subsequent time intervals,
$(\Delta t, 2 \Delta t]$, $(2 \Delta t, 3 \Delta t]$, $\cdots$.
Since ${\cal V} \gg V$, 
the recursion time of an escaped boson to return to 
the original position is extremely long 
(except for rare events whose probability $\to 0$ 
as ${\cal V} \to \infty$.)
Therefore, as long as $J \ll J_{cr}$, 
we can neglect any quantum as well as classical correlations
between transitions of different time intervals.
This allows us to take 
the no-boson state, $| 0^{E} \rangle$, 
as the initial state of the environment for {\em each} time interval
$(\ell \Delta t, (\ell+1) \Delta t]$, where $\ell = 0, 1, 2, \cdots$.
Hence, for every time interval, 
we may use the same formula 
(\ref{rho at Delta t}).
Furthermore, we can neglect the 
$N$ dependencies of $w$ and $J$ under 
our assumption of Eq.\ (\ref{early}).
Therefore, 
\begin{eqnarray}
&& \hat \rho (2 \Delta t)
\nonumber\\
&&
=
w(0;\Delta t) 
\left\{ w(0;\Delta t) {| N, {\bf y} \rangle} {\langle N, {\bf y} |} 
+
w(1;\Delta t) {| N - 1, {\bf y} \rangle} {\langle N - 1, {\bf y} |} 
\right\}
\nonumber\\
&&
\quad +
w(1;\Delta t) 
\left\{
w(0;\Delta t) {| N - 1, {\bf y} \rangle} {\langle N - 1, {\bf y} |} 
+
w(1;\Delta t) {| N - 2, {\bf y} \rangle} {\langle N - 2, {\bf y} |} 
\right\}
\nonumber\\
&&
=
w(0;2 \Delta t) {| N, {\bf y} \rangle} {\langle N, {\bf y} |} 
\nonumber\\
&&
\quad +
w(1;2 \Delta t) {| N - 1, {\bf y} \rangle} {\langle N - 1, {\bf y} |} 
\nonumber\\
&&
\quad +
w(2;2 \Delta t) {| N - 2, {\bf y} \rangle} {\langle N - 2, {\bf y} |}, 
\label{rho at 2 Delta t}\end{eqnarray}
where
\begin{eqnarray}
w(0;2 \Delta t)
&\equiv&
w(0;\Delta t)^2
\nonumber\\
&=&
(1-J \Delta t)^2
\label{w02}\\ 
w(1;2 \Delta t)
&\equiv&
w(0;\Delta t) w(1;\Delta t) 
+
w(1;\Delta t) w(0;\Delta t) 
\nonumber\\
&=&
2 (1-J \Delta t)J \Delta t
\label{w12}\\
w(2;2 \Delta t)
&\equiv&
w(1;\Delta t) w(1;\Delta t) 
\nonumber\\
&=&
J^2 (\Delta t)^2.
\label{w22}\end{eqnarray}
The time evolution 
in the subsequent times can be calculated in a similar manner.
Let
\begin{equation}
t = M \Delta t,
\end{equation}
where $M$ ($<N$) is a positive integer.
We find 
\begin{equation}
\hat \rho(t) 
=
\sum_{m = 0}^{M} w(m;t) {| N - m, {\bf y} \rangle} {\langle N - m, {\bf y} |},
\label{rho at t}\end{equation}
where $w(m; t)$ is the binomial distribution; 
\begin{equation}
w(m; t) 
=
{M \choose m}
(1-J \Delta t)^{M - m}
(J \Delta t)^{m}.
\label{w binomial}\end{equation}
We find from Eq.\ (\ref{rho at t}) that 
$w(m;t)$ is the probability of finding 
$N - m$ bosons in the box at $t$.
From the conservation of the total number of bosons, 
Eq.\ (\ref{conservation of total N}), 
this probability equals the probability
that $m$ bosons have escaped from the box by the time $t$.
Using Eq.\ (\ref{w binomial}), we find
\begin{eqnarray}
\langle N(t) \rangle
&=&
{\rm Tr}  [ \hat \rho(t) \hat N ]
=
\sum_{m = 0}^{M} w(m;t) (N - m)
= 
N - Jt
\\
\langle N^{E}(t) \rangle
&=&
{\rm Tr}  [ \hat \rho(t) \hat N^{E} ]
=
\sum_{m = 0}^{M} w(m;t) m
=
J.
\end{eqnarray}
Since $E_c$ in Eq.\ (\ref{range of Delta t}) 
is of the order of the atomic energy, we can take
$\Delta t$ extremely small such that 
\begin{equation}
M \gg 1
\ \mbox{and} \ 
M \gg Jt,
\label{M is large}\end{equation}
for a finite $t$ that satisfies Eq.\ (\ref{early}).
In this case, 
Eq.\ (\ref{w binomial}) can be approximated by the Poisson 
distribution;
\begin{equation}
w(m; t) 
\approx
K(M, Jt)
e^{- J t}
{
(J t)^{m}
\over
{m}!
}.
\label{w Poisson}\end{equation}
Here, $K$ is the normalization factor,
\begin{equation}
{1 \over K(M,x)}
\equiv
e^{- x}
\sum_{m = 0}^{M} 
{
x^{m}
\over
{m}!
},
\end{equation}
which approaches unity for all $x$ as $M \to \infty$.
For large but finite $M$, we can easily show that  
\begin{equation}
\left|
{1 \over K(M,x)}
-
1
\right|
\sim
{
e^{-x}
\over
\sqrt{2 \pi (M+1)}
}
\left(
{
e x
\over
M + 1
}
\right)^{M+1}.
\label{K is unity}\end{equation}
Therefore, under the condition (\ref{M is large}),  
we can take $K(M, Jt) = 1$
to a very good approximation, and 
we henceforth drop $K$ from Eq.\ (\ref{w Poisson}). 
Furthermore, 
since $w(m,t) \approx 0$ for $m \gg Jt$, 
we may extend the summation of Eq.\ (\ref{rho at t}) to 
$N$.
We thus obtain
\begin{eqnarray}
\hat \rho(t) 
&\approx&
e^{- J t}
\sum_{m = 0}^N
{
(J t)^{m}
\over
{m}!
}
{| N - m, {\bf y} \rangle} {\langle N - m, {\bf y} |}
\nonumber\\
&=&
e^{- J t}
\sum_{m = 0}^N
{
(J t)^{N - m}
\over
(N - m)!
}
{| m, {\bf y} \rangle} {\langle m, {\bf y} |}.
\label{rho Poisson}\end{eqnarray}
Since this final result is valid even at $t=0$ (despite 
our use of the assumption $M \gg 1$), 
it is valid for all $t$ as long as 
\begin{equation}
0 \leq Jt \ll N
\quad \mbox{and} \quad
N \gg 1.
\label{assumption}\end{equation}
Note that the final result (\ref{rho Poisson}) is quite general 
because all the 
details of the box-environment interaction $\hat H^{ES}$ have been 
absorbed in $J$.

The probability $P(m,t)$ of finding $m$ bosons in the box at $t$ is 
evaluated as
\begin{eqnarray}
P(m,t) 
&=&
w(N-m;t)
\nonumber\\
&=&
e^{- J t}
{
(J t)^{N - m}
\over
(N - m)!
}.
\label{sP}\end{eqnarray}
We call this distribution 
the ``shifted Poisson distribution'',
because it is obtained by shifting the center of
the Poisson distribution. 
The expectation values and variances are evaluated as
\begin{eqnarray}
\langle N(t) \rangle
&=&
N - J t
\label{mean N}\\
\langle N^{E}(t) \rangle
&=&
J t
\label{mean NE}\\
\langle \delta N(t)^2 \rangle
&=&
\langle \delta N^{E}(t)^2 \rangle
=
J t.
\label{variance}\end{eqnarray}

\section{Number versus phase}
\label{sec_number vs phase}

\subsection{Cosine and sine operators of interacting many bosons}

Roughly speaking, the conjugate 
observable of the number is the phase.
More precisely, however, 
physical observables are not the phase itself, but the 
cosine and sine of the phase.
Namely, 
any physical measurement of a phase actually measures the cosine or sine 
of the phase \cite{phase}.
In the case of a single-mode boson 
(i.e., a harmonic oscillator of a single degrees of freedom), 
it has been discussed that 
various definitions are possible for the cosine and sine operators 
\cite{mandel}.
This ambiguity does not matter in our case, because
we are treating the case where
the number of bosons is extremely large,
whereas differences among different definitions 
appear only when the number of bosons is small.
On the other hand, the crucial point 
in our case is how to select 
a single ``coordinate'' (dynamical variable) 
with which the phase is defined, 
among a huge degrees of freedom. 
To find such a ``proper coordinate'' is generally very 
difficult in many-body interacting systems.

Fortunately, we find that 
${\hat b_0}$ is the proper coordinate of interacting bosons, 
with which we can successfully define 
the {\em cosine and sine operators of interacting many bosons} by
\begin{eqnarray}
\hat{\cos \phi} 
&\equiv&
{1 \over 2 \sqrt{{\hat b_0^\dagger} {\hat b_0} + 1}} {\hat b_0}
+
{\hat b_0^\dagger} {1 \over 2 \sqrt{{\hat b_0^\dagger} {\hat b_0} + 1}}
\label{cos}\\
\hat{\sin \phi} 
&\equiv&
{1 \over 2 i \sqrt{{\hat b_0^\dagger} {\hat b_0} + 1}} {\hat b_0}
-
{\hat b_0^\dagger} {1 \over 2 i \sqrt{{\hat b_0^\dagger} {\hat b_0} + 1}}.
\label{sin}\end{eqnarray}
These are the same forms as those of
a single harmonic oscillator \cite{mandel}.
In our case, however, there are a huge degrees of freedom
with mutual interactions.
As a result, the Hamiltonian does {\em not} take
the simple bilinear form with respect to ${\hat b_0}$, 
hence the motion of ${\hat b_0} + {\hat b_0^\dagger}$ is not that of 
a harmonic-oscillator coordinate.
Nevertheless, 
many formulas for the single mode case are applicable if 
they are based only on the commutation relations
of a boson operator.
In particular, 
owing to Eq.\ (\ref{GS is vac of bk}), 
we can treat any states that have the form  of
$\sum_m C_m | N-m, {\bf y} \rangle$
as if we were treating a single-mode problem.

It will turn out in the following discussions that 
the above operators give reasonable results for 
the quantum phase of interacting bosons.

\subsection{Number and phase fluctuations of ${| N, {\bf y} \rangle}$}

As we have shown in section \ref{sec_natural},
the ground state of a fixed number of bosons
${| N, {\bf y} \rangle}$ can be represented simply as a number state
if we use a ``natural coordinate'' ${\hat b_0}$.
Note that this state has a finite fluctuation of
$\hat a_0^\dagger \hat a_0$  due to many-body interactions.
Nevertheless, the total number of bosons $\hat N$ 
has a definite value;
\begin{eqnarray}
\langle N \rangle_{N, {\bf y}}
&\equiv&
{\langle N, {\bf y} |} \hat N {| N, {\bf y} \rangle}
=
N,
\\
\langle \delta N^2 \rangle_{N, {\bf y}}
&\equiv&
{\langle N, {\bf y} |} \delta \hat N^2 {| N, {\bf y} \rangle} 
=
0.
\label{N and dN of GS}\end{eqnarray}
On the other hand, using the simple representation (\ref{GS}), 
we can easily show that 
\begin{eqnarray}
\langle \cos \phi \rangle_{N, {\bf y}}
&\equiv&
{\langle N, {\bf y} |} \hat{\cos \phi} {| N, {\bf y} \rangle}
= 0
\label{cos of Ny}\\
\langle \sin \phi \rangle_{N, {\bf y}}
&\equiv&
{\langle N, {\bf y} |} \hat{\sin \phi} {| N, {\bf y} \rangle}
= 0.
\label{sin of Ny}\end{eqnarray}
Therefore, the ground state of a fixed number of bosons 
does not have a definite phase \cite{phase}, 
as expected from the number-phase uncertainty relation, 
Eq.\ (\ref{NPUR}).
It was sometimes argued that 
although $\hat N$ is definite
the fluctuation of $\hat a_0^\dagger \hat a_0$ 
might allow for a definite phase \cite{forster}.
However, our results (\ref{cos of Ny}) and (\ref{sin of Ny}) 
show explicitly that the fluctuation of 
$\hat a_0^\dagger \hat a_0$ does not help to 
develop a definite phase.
Note that this is {\em not} due to our special choice of 
the cosine and sine operators, because 
the same conclusion is obtained also 
when the cosine and sine operators of $\hat a_0$ 
are used instead of Eqs.\ (\ref{cos}) and (\ref{sin}).
We will touch on this point again in section \ref{sec_OP}.

\subsection{Coherent state of interacting bosons}

We define a coherent state of interacting bosons (CSIB) by
\begin{equation}
| \alpha, {\bf y} \rangle
\equiv
e^{-|\alpha|^2/2}
\sum_{m = 0}^\infty
{
\alpha^n
\over
\sqrt{m!}
}
{| m, {\bf y} \rangle},
\label{cs}\end{equation}
which is labeled by ${\bf y}$ and a complex number,
\begin{equation}
\alpha \equiv e^{i \phi} \sqrt{N}.
\end{equation}
The inverse transformation is
\begin{equation}
{| N, {\bf y} \rangle}
=
\int_{-\pi}^{\pi} \frac{d \phi}{2 \pi} 
| \alpha, {\bf y} \rangle.
\label{inverse_tr}\end{equation}
Regarding the number and phase fluctuations, 
we can easily show that 
$| \alpha, {\bf y} \rangle$ has the same properties as 
a coherent state of a single-mode harmonic 
oscillator \cite{mandel}. Namely, 
\begin{eqnarray}
\langle N \rangle_{\alpha, {\bf y}}
&\equiv&
\langle \alpha, {\bf y} | \hat N | \alpha, {\bf y} \rangle
=
|\alpha|^2
=
N,
\\
\langle \delta N^2 \rangle_{\alpha, {\bf y}}
&\equiv&
\langle \alpha, {\bf y} | \delta \hat N^2 | \alpha, {\bf y} \rangle 
=
|\alpha|^2
=
N,
\label{N and dN of CS}\end{eqnarray}
and, for $|\alpha|^2 = N \gg 1$, 
\begin{eqnarray}
\langle \sin \phi \rangle_{\alpha, {\bf y}}
&\equiv&
\langle \alpha, {\bf y} | \hat {\sin \phi} | \alpha, {\bf y} \rangle
\nonumber\\
&=&
[1 - 1 / (8|\alpha|^2) + \cdots ]
\sin \phi,
\\
\langle \delta \sin^2 \phi \rangle_{\alpha, {\bf y}}
&\equiv&
\langle \alpha, {\bf y} | (\delta \hat{\sin \phi})^2 | \alpha, {\bf y} \rangle
\nonumber\\
&=&
(1 / 4|\alpha|^2)
(1 - \sin^2 \phi)
+ \cdots,
\end{eqnarray}
and similar results for the cosine operator.
It is customary to express the results for the sine and 
cosine operators symbolically as
\begin{eqnarray}
\langle \phi \rangle_{\alpha, {\bf y}}
&\approx&
\phi,
\\
\langle \delta \phi^2 \rangle_{\alpha, {\bf y}}
&\approx&
1 / (4|\alpha|^2)
= 1/ (4N).
\end{eqnarray}
Therefore, $| \alpha, {\bf y} \rangle$ is the minimum-uncertainty state 
in the sense that it 
has the {\em minimum} allowable value of the number-phase uncertainty 
product (NPUP) [Eq.\ (\ref{NPUR})];
\begin{equation}
\langle \delta N^2 \rangle_{\alpha, {\bf y}} 
\langle \delta \phi^2 \rangle_{\alpha, {\bf y}}
\approx
1/4.
\label{NPUP_CSIB}\end{equation}
The magnitude of the number fluctuation is conveniently measured with 
the ``Fano factor'' $F$, 
which is defined by  
\begin{equation}
F
\equiv
\langle \delta N^2 \rangle / \langle N \rangle.
\label{Fano}\end{equation}
For $| \alpha, {\bf y} \rangle$, we find
\begin{equation}
F_{\alpha, {\bf y}}
\equiv
{
\langle \delta N^2 \rangle_{\alpha, {\bf y}}
\over
\langle N \rangle_{\alpha, {\bf y}}
}
=
1.
\label{Fano_CSIB}\end{equation}
Therefore, using ${\hat b_0}$, we have successfully constructed a very 
special state of interacting bosons, $| \alpha, {\bf y} \rangle$,
whose Fano factor is {\em exactly} unity, and which 
has the {\em minimum} allowable value of the NPUP.
This should be contrasted with 
Bogoliubov's ground state
$| \alpha_0, {\bf y}^{cl} \rangle^{cl}$, 
for which 
\begin{equation}
F^{cl}
\equiv
{
\langle \delta N^2 \rangle^{cl}
\over
\langle N \rangle^{cl}
}
=
1
+
\frac{
\sum_{{\bf q} \neq {\bf 0}} (\sinh |y_q^{cl}|)^4
}{
|\alpha_0|^2 + \sum_{{\bf q} \neq {\bf 0}} (\sinh |y_q^{cl}|)^2
}
>
1,
\label{Fano cl}\end{equation}
and the NPUP is larger than 1/4.
The CSIB should not be confused with 
$| \alpha_0, {\bf y}^{cl} \rangle^{cl}$.

\subsection{Number-phase squeezed state of interacting bosons}
\label{sec_NPIB}

We define a new state ${| \xi, N, {\bf y} \rangle}$ by
\begin{eqnarray}
{| \xi, N, {\bf y} \rangle} 
&\equiv&
\sqrt{K(N, |\xi|^2)} \ e^{-|\xi|^2/2}
\sum_{n = 0}^N
{
\xi^{*(N - n)}
\over
\sqrt{(N - n)!}
}
{| m, {\bf y} \rangle} 
\\
&=&
\sqrt{K(N, |\xi|^2)} \ e^{-|\xi|^2/2}
\sum_{n = 0}^N
{
\xi^{*(N - n)}
\over
\sqrt{(N - n)! n!}
}
({\hat b_0}^\dagger)^n
{| 0, {\bf y} \rangle},
\label{NPIB}\end{eqnarray}
which is labeled by ${\bf y}$ and a complex number,
\begin{equation}
\xi 
\equiv
e^{i \phi} |\xi|.
\end{equation}
We henceforth assume that 
\begin{equation}
|\xi|^2 \ll N
\quad \mbox{and} \quad
N \gg 1,
\label{assumption_xi}\end{equation}
which allows us to set $K(N, |\xi|^2)=1$ to a very good approximation.

The probability $P(m)$ of finding $m$ bosons for the state ${| \xi, N, {\bf y} \rangle}$ obeys
the shifted Poisson distribution [{\it cf.} Eq.\ (\ref{sP})],
\begin{equation}
P(m) 
=
e^{- |\xi|^2}
{
|\xi|^{2(N - m)}
\over
(N - m)!
}.
\label{sPm}\end{equation}
The number fluctuation and the Fano factor are evaluated as
\begin{eqnarray}
\langle N \rangle_{\xi, N, {\bf y}}
&\equiv&
{\langle \xi, N, {\bf y} |} \hat N {| \xi, N, {\bf y} \rangle}
=
N - |\xi|^2,
\\
\langle \delta N^2 \rangle_{\xi, N, {\bf y}}
&\equiv&
{\langle \xi, N, {\bf y} |} \delta \hat N^2 {| \xi, N, {\bf y} \rangle}
=
|\xi|^2,
\\
F_{\xi, N, {\bf y}}
&\equiv&
{
\langle \delta N^2 \rangle_{\xi, N, {\bf y}}
\over
\langle N \rangle_{\xi, N, {\bf y}}
}
=
{
|\xi|^2
\over
N - |\xi|^2
}
\approx
{
|\xi|^2
\over
N
}
\ll 1.
\end{eqnarray}
As compared with Eqs.\ (\ref{N and dN of CS}) and 
(\ref{Fano_CSIB}), we observe that 
the state ${| \xi, N, {\bf y} \rangle}$ has a very narrow distribution of the boson number.
On the other hand, 
${| \xi, N, {\bf y} \rangle}$ has a well-defined phase \cite{phase} when 
\begin{equation}
1 \ll |\xi|^2 \ll N.
\label{xi N}\end{equation}
In fact, under this condition we can easily show that 
\begin{eqnarray}
\langle \sin \phi \rangle_{\xi, N, {\bf y}}
&\equiv&
{\langle \xi, N, {\bf y} |} \hat {\sin \phi} {| \xi, N, {\bf y} \rangle}
\nonumber\\
&=&
\left[1 - 1 /(8|\xi|^2) + \cdots \right]
\sin \phi
\\
\langle \delta \sin^2 \phi \rangle_{\xi, N, {\bf y}}
&\equiv&
{\langle \xi, N, {\bf y} |} (\delta \hat{\sin \phi})^2 {| \xi, N, {\bf y} \rangle}
\nonumber\\
&=&
(1 / 4|\xi|^2)
(1 - \sin^2 \phi)
+ \cdots,
\end{eqnarray}
and similar results for the cosine operator.
As in the case of the CSIB, 
we may express these results symbolically as
\begin{eqnarray}
\langle \phi \rangle_{\xi, N, {\bf y}}
&\approx&
\phi
\\
\langle \delta \phi^2 \rangle_{\xi, N, {\bf y}}
&\approx&
1 / (4|\xi|^2).
\end{eqnarray}
Therefore, 
just as $| \alpha, {\bf y} \rangle$ does,
${| \xi, N, {\bf y} \rangle}$ has the minimum value of 
the NPUP;
\begin{equation}
\langle \delta N^2 \rangle_{\xi, N, {\bf y}}
\langle \delta \phi^2 \rangle_{\xi, N, {\bf y}}
\approx 1/4
\quad \mbox{(for $1 \ll |\xi|^2 \ll N$)}.
\label{NPUP_NPIB}\end{equation}
Since each component of the product satisfies
$
\langle \delta \hat N^2 \rangle_{\xi, N, {\bf y}}
\ll
\langle \delta \hat N^2 \rangle_{\alpha,{\bf y}}
$
and 
$
\langle \delta \phi^2 \rangle_{\xi, N, {\bf y}}
\gg
\langle \delta \phi^2 \rangle_{\alpha,{\bf y}}
$, 
${| \xi, N, {\bf y} \rangle}$ 
is obtained by 
``squeezing'' $| \alpha, {\bf y} \rangle$
in the direction of $\hat N$,
while keeping the NPUP minimum.
({\it cf.} The conventional squeezed state 
has a larger NPUP \cite{mandel}.)
We thus call 
${| \xi, N, {\bf y} \rangle}$ the 
``number-phase squeezed state of interacting bosons'' (NPIB).
 
\subsection{Phase-randomized mixture of 
number-phase squeezed states of interacting bosons}
\label{sec_PRM}

We now take
\begin{equation}
\xi 
=
e^{i \phi} \sqrt{Jt} 
\equiv
\xi(t).
\end{equation}
That is, 
$
|\xi|^2 = J t
$.
Then, inequalities (\ref{assumption_xi}) are satisfied because of 
our assumption (\ref{assumption}).
We can show by explicit calculation that 
Eq.\ (\ref{rho Poisson}) can be rewritten as
\begin{equation}
\hat \rho(t) 
=
\int_{- \pi}^{\pi} {d \phi \over 2 \pi}
{| e^{i \phi} \sqrt{Jt}, N, {\bf y} \rangle} {\langle e^{i \phi} \sqrt{Jt}, N, {\bf y} |}.
\label{rho random phase}\end{equation}
Therefore, 
the boson state in the box can be viewed 
either as the shifted Poissonian mixture, Eq.\ (\ref{rho Poisson}),  
of NSIBs, 
or as the phase-randomized mixture (PRM), Eq.\ (\ref{rho random phase}), 
of NPIBs.
Both representations are simply described in terms of $\hat b_0$. 
In contrast, the same $\hat \rho(t)$ would be described in an 
very complicated manner in terms of bare operators.

We have thus obtained {\em double pictures} (or {\em representations}), 
Eqs.\ (\ref{rho Poisson}) and (\ref{rho random phase}),
for the {\em same} physical state \cite{double,double_pic}.
According to the former picture, 
the state of the box is one of NSIBs, 
for which the number of bosons is definite (but unknown), whereas
the phase is completely indefinite.
According to the latter picture, on the other hand, 
the state is one of NPIBs, 
for which the number of bosons has a finite fluctuation 
$\langle \delta N^2 \rangle \approx Jt$, whereas
the phase is almost definite \cite{phase} (but unknown), 
$\langle \delta \phi^2 \rangle \approx 1/(4 Jt)$.
What allows these double pictures is the superposition principle
\cite{double_pic}.

In addition to Eqs.\ (\ref{rho Poisson}) and (\ref{rho random phase}),
there are many other ways to express $\hat \rho (t)$ as 
different mixtures.
Among them, Eq.\ (\ref{rho Poisson}) is the form in which 
{\em each} element of the mixture has the smallest value of 
the number fluctuation, 
whereas in Eq.\ (\ref{rho random phase}) 
{\em each} element of the mixture has the smallest value of 
the phase fluctuation.
Therefore, the latter representation is particularly convenient for 
discussing physical properties that are related to the phase,
as will be shown in sections \ref{sec_pm} and \ref{sec_OP}.

\subsection{Origin of the direction of the time evolution}

As the time evolves, 
the number of bosons decreases as
\begin{equation}
\langle N(t) \rangle
=
N - Jt.
\label{Nt}\end{equation}
As a result, the energy of the bosons in the box decreases with $t$.
For example, for each element of Eq.\ (\ref{rho random phase}),
\begin{equation}
{\langle e^{i \phi} \sqrt{Jt}, N, {\bf y} |} \hat H {| e^{i \phi} \sqrt{Jt}, N, {\bf y} \rangle}
<
{\langle N, {\bf y} |} \hat H {| N, {\bf y} \rangle}
\quad \mbox{for $t>0$}.
\end{equation}
However, we note that this energy difference is just 
a consequence of the difference in $\langle N \rangle$.
Namely, 
${| N, {\bf y} \rangle}$ and the NPIB have the same energy 
if they have the same value of $\langle N \rangle$;
\begin{equation}
\langle e^{i \phi} \sqrt{Jt}, N+Jt, {\bf y} |
\hat H
| e^{i \phi} \sqrt{Jt}, N+Jt, {\bf y} \rangle
\approx
{\langle N, {\bf y} |} \hat H {| N, {\bf y} \rangle}.
\label{same_energy}\end{equation}
We can therefore conclude that 
the direction of the time evolution, 
from ${| N, {\bf y} \rangle}$ to the PRM of ${| \xi, N, {\bf y} \rangle}$, is
{\em not} determined by an energy difference.
Hence, it must be due to difference in the 
nature of the wavefunctions.
The study of such a nature, however, needs many pages of analysis,
which is beyond the scope 
of this paper, 
thus will be described elsewhere \cite{unpublished}.

\section{Action of measurement or its equivalence}
\label{sec_action of meas.}

In the previous section, 
we have obtained the double pictures, 
Eqs.\ (\ref{rho Poisson}) and (\ref{rho random phase}).
Depending on the physical situation, 
either picture is convenient \cite{double_pic}.
To explain this point, we discuss two examples in this section.

\subsection{Number measurement}

Suppose that one measures $N$ 
of the boson system whose density operator is given by 
Eq.\ (\ref{rho Poisson}), or, 
equivalently, by Eq.\ (\ref{rho random phase}).
In this case, the former expression is convenient.
In fact, 
if the measurement error
\begin{equation}
\delta N_{err} 
<
\sqrt{\langle \delta N(t)^2 \rangle}
=
\sqrt{Jt},
\end{equation}
and if the measurement is of the first kind \cite{SF}, 
then the action of the measurement is 
to narrower the number distribution as small as
$\delta N_{err}$, of 
the rhs of Eq.\ (\ref{rho Poisson}).
That is, 
the density operator immediately after the measurement 
is generally given by \cite{mtheory}
\begin{equation}
\hat \rho_{\bar{N}}(t) 
\equiv
\sum_{m = 0}^N
W(m - \bar{N})
{| m, {\bf y} \rangle} {\langle m, {\bf y} |}.
\label{rho number meas}\end{equation}
Here, 
$\bar{N}$ is the value of $\phi$ obtained by the measurement, and 
$W$ is a smooth function which has
the following properties:
\begin{eqnarray}
&& 
W(m - \bar{N}) \geq 0,
\\
&&
W(m - \bar{N}) \approx 0
\quad \mbox{for} \quad
|m - \bar{N}| \gtrsim \delta N_{err},
\\
&&
\sum_{m = 0}^N
W(m - \bar{N})
= 1.
\end{eqnarray}
The detailed form of $W$ depends on the detailed structures of the 
measuring apparatus, and thus is of no interest here.
In an ideal case where $\delta N_{err} \to 0$, 
$W$ becomes Kronecker's delta, and
\begin{equation}
\hat \rho_{\bar{N}}(t) 
\to
{| \bar{N}, {\bf y} \rangle} {\langle \bar{N}, {\bf y} |}.
\end{equation}
On the other hand, if the measurement is rather inaccurate
in such a way that  
\begin{equation}
\delta N_{err} 
>
\sqrt{\langle \delta N(t)^2 \rangle}
=
\sqrt{Jt},
\end{equation}
then almost no change of $\hat \rho$ is induced by the measurement, 
if it is of the first kind \cite{mtheory}.

\subsection{Phase measurement}
\label{sec_pm}

Suppose that one measures $\phi$ 
of the boson system whose density operator is 
Eq.\ (\ref{rho Poisson}), or, 
equivalently, Eq.\ (\ref{rho random phase}).
In this case, the latter representation is convenient.
In fact, 
if the measurement error
\begin{equation}
\delta \phi_{err} 
>
\sqrt{\langle \delta \phi^2 \rangle_{\xi, N, {\bf y}}}
=
{1
\over
2 |\xi(t)|
}
=
{1
\over
2 \sqrt{Jt}
},
\end{equation}
and if 
the measurement is performed in such a way that
the backaction of the measurement is minimum, 
then the action of the measurement is 
just to find (or, get to know) the ``true'' value of $\phi$,
to the accuracy of $\delta \phi_{err}$,
among many possibilities in the rhs side of 
Eq.\ (\ref{rho random phase}).
Therefore, 
the density operator just after the measurement is generally given 
by \cite{mtheory}
\begin{equation}
\hat \rho_{\bar{\phi}}(t) 
\equiv
\int_{- \pi}^{\pi} {d \phi \over 2 \pi}
D(\phi - \bar{\phi}) {| e^{i \phi} \sqrt{Jt}, N, {\bf y} \rangle} {\langle e^{i \phi} \sqrt{Jt}, N, {\bf y} |}.
\label{rho phase meas}\end{equation}
Here, 
$\bar{\phi}$ is the value of $\phi$ obtained by the measurement, and 
$D(\phi - \bar{\phi})$ is a smooth function which has
the following properties:
\begin{eqnarray}
&& 
D(\phi - \bar{\phi}) \geq 0,
\\
&&
D(\phi - \bar{\phi}) \approx 0
\quad \mbox{for} \quad
|\phi - \bar{\phi}| \gtrsim \delta \phi_{err},
\\
&&
\int_{- \pi}^{\pi} {d \phi \over 2 \pi}
D(\phi - \bar{\phi}) 
= 1.
\end{eqnarray}
The detailed form of $D$ depends on the detailed structures of the 
measuring apparatus, and thus is of no interest here.

On the other hand, if the measurement is very accurate 
in such a way that
\begin{equation}
\delta \phi_{err} 
<
\sqrt{\langle \delta \phi^2 \rangle_{\xi, N, {\bf y}}}
=
{1
\over
2 |\xi(t)|
}
=
{1
\over
2 \sqrt{Jt}
},
\end{equation}
then $\hat \rho$ will ``collapse'' into 
another state whose phase fluctuation is less than 
$\langle \delta \phi^2 \rangle_{\xi, N, {\bf y}}$.
However, such an accurate measurement is practically difficult when 
$Jt \gg 1$.
Therefore, in most experiments
we may take 
Eq.\ (\ref{rho phase meas}) for 
the density operator after the measurement, 
if the measurement is performed in such a way that
the backaction of the measurement is minimum.

\begin{figure}[h]
\begin{center}
\epsfile{file=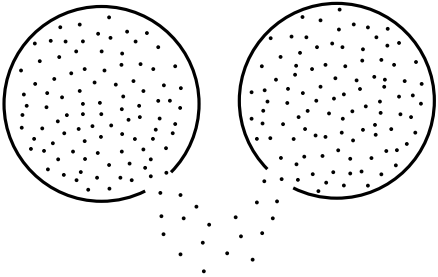,scale=0.4}
\end{center}
\caption{Bosons are confined independently in two boxes.
The number of bosons in each box is fixed for $t<0$.
If holes are made in the boxes at $t>0$, then the leakage fluxes
are induced, which exhibit interference.
}
\label{two boxes}
\end{figure}

For example, suppose that 
one prepares bosons {\em independently}
in two boxes [Fig.\ \ref{two boxes}],
where the number of bosons in each box is fixed for $t<0$.
At $t=0$ a small hole is made in each box, and 
small fluxes of bosons escape from the boxes. 
Since $\hat \rho$ of each box evolves as 
Eq.\ (\ref{rho random phase}), 
it is clear that 
these fluxes can interfere 
{\em at each experimental run}
(as in the cases of non-interacting bosons \cite{theory1,theory2,theory3} 
and two-mode bosons \cite{RW})
as the interference between 
two NPIBs of two boxes.
From the interference pattern, 
one can measure the relative phase $\bar \phi$ of two condensates.
After the measurement, 
$\hat \rho$ of each box would take 
the form of eq.\ (\ref{rho phase meas}).
Such an experiment would be possible 
by modifying the experiment of Ref.\ \cite{andrews}
in such a way that $J$ becomes small enough.

\section{Order parameter}
\label{sec_OP}

The order parameter of BEC
can be defined in various ways.
For the ground state 
$| \alpha, {\bf y}^{cl} \rangle^{cl}$
of the semiclassical Hamiltonian $\hat H^{cl}$, 
different definitions give the same result.
However, this is not the case for 
${| N, {\bf y} \rangle}$ 
and ${| \xi, N, {\bf y} \rangle}$.
We explore these points in this section.

\subsection{Off-diagonal long-range order}

We first consider the two-point 
correlation function defined by
\begin{equation}
\Upsilon({\bf r}_1, {\bf r}_2)
\equiv
{\rm Tr}  [ \hat \rho
\hat \psi^\dagger({\bf r}_1) \hat \psi({\bf r}_2)].
\label{Upsilon}\end{equation}
The system is said to possess the 
off-diagonal long-range order (ODLRO) if \cite{penrose,yang,odlro}
\begin{equation}
\lim_{|{\bf r}_1 - {\bf r}_2| \to \infty}
\Upsilon({\bf r}_1, {\bf r}_2)
\neq
0.
\label{ODLRO}\end{equation}
This limiting value cannot be finite 
without the condensation of a macroscopic number of bosons.
(Without the condensation, we simply has 
$
\lim_{|{\bf r}_1 - {\bf r}_2| \to \infty}
\Upsilon({\bf r}_1, {\bf r}_2)
=0
$
for the ground state and for any finite excitations.)
In this sense, 
Eq.\ (\ref{ODLRO}) is a criterion of the condensation.

If the system possesses the ODLRO, 
it is customary to define 
the order parameter $\Xi$ by 
the asymptotic form of $\Upsilon$ as
\begin{equation}
\Upsilon({\bf r}_1, {\bf r}_2)
\sim
\Xi^*({\bf r}_1) \Xi({\bf r}_2).
\label{Xi}\end{equation}
According to this definition, 
we obtain the same results
for {\em all} of 
${| N, {\bf y} \rangle}$, ${| \xi, N, {\bf y} \rangle}$, and 
$| \alpha, {\bf y} \rangle$,
where $\xi = e^{i \phi} \sqrt{Jt}$,
$Jt \ll N$ and $\alpha = e^{i \phi} \sqrt{N}$.
Namely, using Eqs.\ (\ref{property of psi'}) and
(\ref{formula of decomposition}), we find 
\begin{equation}
\lim_{|{\bf r}_1 - {\bf r}_2| \to \infty}
\Upsilon({\bf r}_1, {\bf r}_2)
=
n_0,
\ \mbox{hence} \ \Xi = \sqrt{n_0} e^{i \varphi},
\label{ODLRO of all states}\end{equation}
for {\em all} of these states.
Therefore, neither the ODLRO nor $\Xi$
is able to distinguish between these states.

\subsection{Definition as a matrix element}

As an order parameter of 
the state for which $N$ is exactly fixed, 
Ref.\ \cite{LP} uses the ``wavefunction of
the condensate'' $\Xi$, as defined by Eq.\ (\ref{wf of condensate}).
It is clear that this definition is just a special case of 
that of the previous section.
In fact, for ${| N, {\bf y} \rangle}$ Eq.\ (\ref{wf of condensate}) yields
\begin{equation}
\Xi 
=
{\langle N - 1, {\bf y} |} \hat \psi {| N, {\bf y} \rangle}
= \sqrt{n_0} \ e^{i \varphi},
\end{equation}
in agreement with Eq.\ (\ref{ODLRO of all states}).

\subsection{Definition as the expectation value of $\hat \psi$}

Another definition 
of the order parameter is 
the expectation value of $\hat \psi$, which we denote by $\Psi$;
\begin{equation}
\Psi ({\bf r}) \equiv \langle \hat \psi({\bf r}) \rangle.
\end{equation}
According to this definition, 
the ground state ${| N, {\bf y} \rangle}$ 
of a fixed number of bosons
does not have a finite order parameter;
\begin{equation}
\Psi
=
{\langle N, {\bf y} |} \hat \psi({\bf r}) {| N, {\bf y} \rangle}
=
0.
\label{Psi_is_zero}\end{equation}
This result is rather trivial because $\hat \psi$ 
alters $N$ exactly by one.
On the other hand, it was sometimes conjectured
in the literature \cite{forster} that
the expectation value of the bare operator $\hat a_0$ 
might be finite 
$
{\langle N, {\bf y} |} \hat a_0 {| N, {\bf y} \rangle} \neq 0
$
in the presence of many-body interactions
because the number of bosons in the bare state of ${\bf k} = 0$ 
fluctuates due to the many-body scatterings.
However, this conjecture is wrong because 
by integrating Eq.\ (\ref{Psi_is_zero}) over ${\bf r}$ 
we obtain
\begin{equation}
{\langle N, {\bf y} |} \hat a_0 {| N, {\bf y} \rangle} = 0.
\end{equation}
That is, although $\hat a_0^\dagger \hat a_0$ fluctuates 
in the state ${| N, {\bf y} \rangle}$ it does not lead to a finite 
${\langle N, {\bf y} |} \hat a_0 {| N, {\bf y} \rangle}$.

In our gedanken experiment,
$\hat \rho(t)$ evolves as
Eq.\ (\ref{rho Poisson}), or, equivalently, 
as Eq.\ (\ref{rho random phase}).
For this mixed ensemble, 
\begin{equation}
\Psi = {\rm Tr}[\hat \rho(t) \hat \psi] = 0.
\end{equation}
This is the {\em average over all elements} 
in the mixed ensemble, 
and corresponds to the {\em average over many experimental runs}.
On the other hand, 
$\Psi$ of {\em each element}, 
which corresponds to {\em a possible 
result for a single experimental run},
is different between 
the two expressions, Eqs.\ (\ref{rho Poisson}) and (\ref{rho random phase}).
That is, for each element of the mixtures, 
$\Psi = 0$ for Eq.\ (\ref{rho Poisson}) 
because of Eq.\ (\ref{Psi_is_zero}), 
whereas for Eq.\ (\ref{rho random phase})
\begin{equation}
\Psi
=
{\langle e^{i \phi} \sqrt{Jt}, N, {\bf y} |} \hat \psi {| e^{i \phi} \sqrt{Jt}, N, {\bf y} \rangle}
\label{maxPsi}\end{equation} 
can be finite. 
As discussed in sections \ref{sec_NPIB} and \ref{sec_PRM}, 
Eq.\ (\ref{rho random phase}) 
is the form in which 
{\em each} element of the mixture has the smallest value of 
the phase fluctuation.
This indicates that 
each element of the mixture possesses 
the most definite (non-fluctuating) value 
of $\Psi$
when we take the representation (\ref{rho random phase}), 
among many representations of the same $\hat \rho(t)$.
We are most interested in this case, because 
$\Psi$ is usually taken as a macroscopic order parameter, 
which has a definite value obeying the Ginzburg-Landau equation.

\begin{figure}[h]
\begin{center}
\epsfile{file=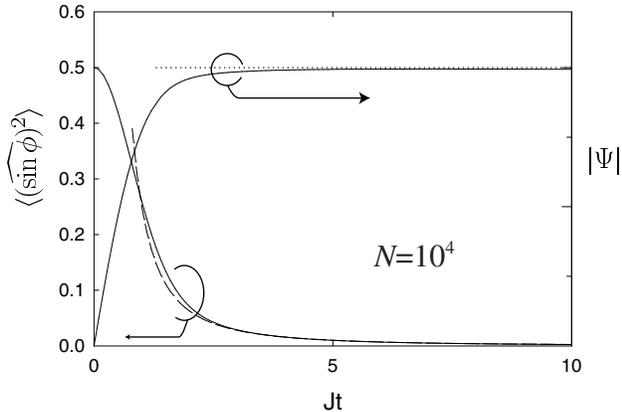,scale=0.5}
\end{center}
\caption{
Left scale: $\langle (\hat{\sin} \phi)^2 \rangle$ 
of the $\phi=0$ element of Eq.\ (\protect\ref{rho random phase}).
The dashed line represents $1/(4Jt)$.
Right scale: $|\Psi|$ defined by Eq.\ (\protect\ref{maxPsi}).
The dotted line denotes $|\Psi|=\protect\sqrt{n_0}$.
Both $\langle (\hat{\sin} \phi)^2 \rangle$ 
and $|\Psi|$
are plotted against $Jt$, the number of escaped bosons.
}
\label{OP}
\end{figure}
Figure \ref{OP} plots 
$|\Psi|$ of Eq.\ (\ref{maxPsi})
as a function of $Jt$ ($=$ the average number of escaped bosons).
We find that $|\Psi|$ grows very rapidly,
until it attains a constant value\cite{decrease},  
\begin{equation}
\Psi
=
{\langle e^{i \phi} \sqrt{Jt}, N, {\bf y} |} \hat \psi {| e^{i \phi} \sqrt{Jt}, N, {\bf y} \rangle}
\to
e^{i (\phi + \varphi)} \sqrt{n_0} 
\quad \mbox{for $Jt \gtrsim 2$}.
\end{equation} 
This value equals 
$\Psi$ of 
$| \alpha, {\bf y} \rangle$ with $\alpha = e^{i \phi} \sqrt{N}$;
\begin{equation}
\Psi
=
\langle \alpha, {\bf y} | \hat \psi | \alpha, {\bf y} \rangle
=
e^{i \varphi} \sqrt{n_0 \over n V} \alpha
=
e^{i (\phi + \varphi)} \sqrt{n_0}.
\end{equation}
Note here that $|\Psi|$ of $| \alpha, {\bf y} \rangle$ is 
renormalized by the factor $\sqrt{|Z|} = \sqrt{n_0/n}$ because of the
many-body interactions.

We have also plotted $\langle (\hat{\sin} \phi)^2 \rangle$ 
of the $\phi=0$ element of Eq.\ (\ref{rho random phase})
in Fig.\ \ref{OP}. 
This is a measure of
the phase fluctuation of the $\phi=0$ element.
Because of the rotational symmetry with respect to $\phi$, 
we can regard it as a measure of
the phase fluctuation of every element.
We find that 
$\langle (\hat{\sin} \phi)^2 \rangle$ decreases rapidly 
as $Jt$ is increased, until 
it behaves as 
\begin{equation}
\langle (\hat{\sin} \phi)^2 \rangle
\approx
1/(4Jt),
\end{equation}
for $Jt \gtrsim$ 3. 
Therefore, 
{\em after the leakage of only two or three bosons,
${| e^{i \phi} \sqrt{Jt}, N, {\bf y} \rangle}$ acquires the 
full, stable and definite (non-fluctuating) values of $\Psi$ and $\phi$}, 
and the {\em gauge symmetry is broken} in this sense.
One might expect that 
$\langle \delta N^2 \rangle$ of the order of $\langle N \rangle$ 
would be necessary to achieve such stable $\Psi$ and $\phi$
because $\langle \delta N^2 \rangle = \langle N \rangle$ for 
a CSIB.
Our result shows that this expectation is wrong, because 
$\Psi$ and $\phi$ already become stable when $Jt \sim 2$, for which 
$\langle \delta N^2 \rangle = Jt \ll \langle N \rangle$.
Practically, it seems rather difficult to 
fix $N$ to such high accuracy that $\delta N \lesssim 2$.
In such a case, 
$\delta N$ would be larger than 2 from the beginning,
and each element of the mixture
has the full and stable values of $\Psi$ and $\phi$ from the beginning.

We finally make a remark on the evolution at later times, 
whereas we have only 
considered the early time stage for which $Jt \ll N$.
It is clear that 
the system eventually approaches the equilibrium state.
However, a question is; what is the state 
after the early stage, but before the 
system reaches the equilibrium?
It is expected that the state would be 
some coherent state.
We can show that this is indeed the case \cite{unpublished}: 
as $t \to \infty$,  
$\hat \rho$ eventually approaches  
the PRM of $| \alpha, {\bf y} \rangle$,  
in which  
$|\alpha|^2 = \langle N(t) \rangle$ ($<N$) \cite{decrease}
and ${\bf y}$ is given by Eqs.\ 
(\ref{cosh y cl}), (\ref{sinh y cl}) and (\ref{yq})
with $n = \langle N(t) \rangle/V$ \cite{decrease}.
To show this, we must extend the theory of section \ref{sec_evolution}.
This is beyond the scope 
of this paper, and thus will be described elsewhere \cite{unpublished}.

The summary of the present paper has been given in section \ref{intro}.

\acknowledgments{
Helpful discussions with 
M.\ Ueda, K.\ Fujikawa, H.\ Fukuyama, 
T.\ Kimura and T.\ Minoguchi
are acknowledged.
The authors also thank M.\ D.\ Girardeau for informing them of
Ref.\ \cite{GA}.
}

%


\begin{references}

\bibitem[*]{shmz} Electronic address: shmz@ASone.c.u-tokyo.ac.jp

\bibitem[**]{inoue} Electronic address: inoue@ASone.c.u-tokyo.ac.jp

\bibitem{bec93}
For reviews, see, e.g., papers in 
A.\ Griffin {\it et al.} (eds.), 
Bose-Einstein Condensation, Cambridge, New York, 1995.

\bibitem{He4}
P.\ Sokol, p.\ 51 of Ref. \cite{bec93}.
 
\bibitem{BECexciton}
 E. Fortin, S. Fafard, and A. Mysyrowicz,
 Phys. Rev. Lett. {\bf 70}, 3861 (1993);
%
J. P. Wolfe, J. L. Lin and D. W. Snoke, 
p.\ 281 of Ref. \cite{bec93};
%
 L.V. Butov {\it et al.}, 
 {\it ibid}, {\bf 73}, 304 (1994)



\bibitem{BECatom1}
M.\ H.\ Anderson {\it et al.}:
Science {\bf 269} (1995) 198.

\bibitem{BECatom2}
C.\ C.\ Bradley {\it et al.}:
Phys. Rev. Lett. {\bf 75} (1995) 1687.

\bibitem{BECatom3}
K.\ B.\ Davis {\it et al.}:
Phys.\ Rev.\ Lett.\ {\bf 75} (1995) 3969.


\bibitem{Noz}
P.\ Nozi\`eres,
p.\ 15 of Ref.\ \cite{bec93}.


\bibitem{popov}
V.\ N.\ Popov,
{\it Functional integrals and collective excitations},
(Cambridge, New York, 1987).


\bibitem{He4bubble}
E.\ G.\ Syskakis, F.\ Pobell and H. Ullmaier, 
Phys. Rev. Lett. {\bf 55} (1985) 2964.

\bibitem{GA}
M.\ Girardeau and R.\ Arnowitt,
Phys. Rev. {\bf 113}, 755 (1959);
M.\ D.\ Girardeau, 
Phys. Rev. {\bf A58}, 775 (1998).

\bibitem{theory1}
J.\ Javanainen and S.\ M.\ Yoo,  
Phys. Rev. Lett. {\bf 76} (1996) 161.

\bibitem{theory2}
M.\ Naraschewski {\it et al.}, 
Phys.\ Rev.\ A {\bf 54} (1996) 2185.

\bibitem{theory3}
Y.\ Castin and J.\ Dalibard, 
Phys.\ Rev.\ A {\bf 55} (1997) 4330.

\bibitem{gold}
N.\ Goldenfeld,
{\it Lectures on Phase Transitions and the 
Renormalization Group}
(Addison-Wesley, New York, 1992).

\bibitem{LP} 
E.M.\ Lifshitz and L.P.\ Pitaevskii,
{\it Statistical Physics Part 2} 
(Pergamon, New York, 1980), sec.\ 26.


\bibitem{root}
Throughout this paper, we use $\sqrt{\ \ }$ to denote 
the positive square root.


\bibitem{gardiner}
C.W.\ Gardiner, 
Phys.\ Rev.\ A {\bf 56} (1997) 1414.

\bibitem{castin}
Y. Castin and R. Dum, 
Phys.\ Rev.\ A {\bf 57} (1998) 3008.


\bibitem{Leg}
A.J.\ Leggett,
p.\ 452 of Ref.\ \cite{bec93}.


\bibitem{phase}
When we talk about the phase, 
we of course mean the phase {\em relative to}
some reference.



\bibitem{note_vac}
Actually, 
when $V$ is finite, 
we can easily show that 
$
{| 0, {\bf y} \rangle}
=
| 0 \rangle
$.
This equality, however, becomes meaningless when $V \to \infty$ 
and/or when $\hat a_0$ were replaced with a c-number.
Anyway,we do not use this equality in the present paper.


\bibitem{hermitian}
We can add its conjugate (creation operator) if 
we want to make the ``coordinate'' hermitian.

\bibitem{RW}
J.\ Ruostekoski and D.\ F.\ Walls, 
Phys.\ Rev.\ A {\bf 58} (1998) R50.


\bibitem{Jcr}
Since $J_{cr}$ strongly depends on the structures of 
the box and the hole or walls, 
$J_{cr}$ should be determined experimentally.

\bibitem{mandel}
See, e.g., L. Mandel and E. Wolf, \it
Optical Coherence and Quantum Optics \rm
(Cambridge Univ.\ Press, 1995).


\bibitem{forster}
See, e.g., D.\ Forster, 
{\it Hydrodynamic Fluctuations, Broken Symmetry, and
Correlation Functions}
(Benjamin, London, 1975), section 10.3.

\bibitem{double}
Similar double pictures are well-known 
for free bosons,  
as the equivalence of the Poissonian mixture of number states 
and the phase-randomized mixture of coherent states 
\cite{theory1,theory2,theory3}.


\bibitem{double_pic}
The simplest example of the double (or more) picture is 
a mixed state of a single spin.
If the quantum state of a single spin-1/2 system is 
either the up-spin state
$| +z \rangle$ or the down-spin state $| -z \rangle$, 
and if we have no information on which state is realized
[in other words, the $z$  component of the spin is completely 
undefined], 
then the density matrix is, 
\begin{equation}
\hat \rho =
\frac{1}{2}| +z \rangle \langle +z | 
+ \frac{1}{2}| -z \rangle \langle -z |.
\label{rho1}\end{equation}
On the other hand, 
if the direction of the spin is either 
in the $+x$ direction 
$| +x \rangle$ or 
in the $-x$ direction 
$| -x \rangle$, 
and if we have no information on which state is realized, 
then the density matrix is, 
\begin{equation}
\hat \rho =
\frac{1}{2}| +x \rangle \langle +x | 
+ \frac{1}{2}| -x \rangle \langle -x |.
\label{rho2}\end{equation}
It is easy to verify that Eqs.\ (\ref{rho1}) and (\ref{rho2}) are
identical.
Therefore, the same density matrix can be interpreted in two 
ways; 
the $z$ component of the spin is definite, but unknown 
[Eq.\ (\ref{rho1})];
the $x$ component of the spin is definite, but unknown
[Eq.\ (\ref{rho2})].
Moreover, $\hat \rho$ has many other representations, 
e.g., 
\begin{equation}
\hat \rho =
\frac{1}{2}| +y \rangle \langle +y | 
+ \frac{1}{2}| -y \rangle \langle -y |.
\label{rho3}\end{equation}
As demonstrated by this simple example, 
a mixed state of a quantum system generally has many 
representations, hence has many interpretations or pictures.
[This is a result of the superposition principle.]
Some picture is convenient depending on the physical situation.
For example, Eq.\ (\ref{rho2}) is most convenient to 
analyze the case where the $x$ component of the spin is measured.

\bibitem{unpublished}
A. Shimizu et al., 
unpublished.



\bibitem{SF}
See , e.g., A. Shimizu and K. Fujita, 
{\it Quantum Control and Measurement}
[H. Ezawa and Y. Murayama, eds., North-Holland, Amsterdam, 1993]
p.\ 191.
(quant-ph/9804026)


\bibitem{mtheory}
Since the pre-measurement state
[Eq.\ (\ref{rho Poisson}), or, equivalently, Eq.\ (\ref{rho random phase})]
is a mixed state, 
the post-measurement state is also a mixed state 
if the measurement error is finite: 
the post-measurement state can be pure only after an error-less
measurement.
This may be understood by considering the measurement error and 
the post-measurement state as functions of the coupling strength $G$ 
between the measured system and the measuring apparatus:
As $G$ is decreased, the measurement error increases, whereas 
the backaction on the measured system decreases. In the limit of 
$G \to 0$, the measurement error reaches 100 \% 
(i.e., no information can be obtained by such a silly measurement)  and 
the density operator of the measured system does not 
change at all (hence remains a mixed state).
As $G$ is increased, 
on the other hand, 
the measurement error decreases, whereas
the post-measurement state approaches a pure state.



\bibitem{andrews}
M.\ R.\ Andrews {\it et al.}, 
Science {\bf 275} (1997) 637.
This type of experiment is approximately of the first kind
if the leakage fluxes are small enough.


\bibitem{penrose}
O.\ Penrose and L.\ Onsager, 
Phys. Rev. {\bf 104} (1956) 413.

\bibitem{yang}
C.N.\ Yang, Rev.\ Mod.\ Phys. {\bf 34}, 694 (1962).

\bibitem{odlro}
Precisely speaking, 
since we consider the case of {\em finite} volume $V$,
Eqs.\ (\ref{ODLRO}) and (\ref{ODLRO of all states}) should 
be expressed as the following asymptotic behaviors 
as $|{\bf r}_1 - {\bf r}_2| \sim V^{1/3}$;
\begin{eqnarray}
\Upsilon({\bf r}_1, {\bf r}_2)
&\sim& 
C V^0
\quad (C \neq 0),
\\
\Upsilon({\bf r}_1, {\bf r}_2)
&\sim&
n_0 V^0.
\end{eqnarray}


\bibitem{decrease}
This value 
decreases gradually because 
$\langle N(t) \rangle$ ($\propto n_0$) decreases with time.


\end{references}
\end{document}